\documentclass[twocolumn]{aastex62}
\usepackage[utf8]{inputenc}
\usepackage{amsmath}
\usepackage{threeparttable}
\usepackage{enumitem}

\newcommand\ciimath{\operatorname{[CII]}}

\graphicspath{{./}{figures_ApJ/}}

\shorttitle{[CII] in ASPECS}

\begin{document}
\title{The ALMA Spectroscopic Survey in the HUDF: A Search for [CII] Emitters at $6 \leq z \leq 8$}

\correspondingauthor{Bade D. Uzgil}
\email{badeu@astro.caltech.edu}

\author{Bade D. Uzgil}
\affil{California Institute of Technology, 1200 E. California Blvd, Pasadena, CA 91125, USA}
\affil{National Radio Astronomy Observatory, Pete V. Domenici Array Science Center, P.O. Box 0, Socorro, NM 87801, USA}

\author{Pascal A. Oesch}
\affil{Department of Astronomy, University of Geneva, Ch. des Maillettes 51, 1290 Versoix, Switzerland}
\affil{International Associate, Cosmic Dawn Center (DAWN) at the Niels Bohr Institute, University of Copenhagen and DTU-Space, Technical University of Denmark, Copenhagen, Denmark}

\author{Fabian Walter}
\affil{Max-Planck-Institut f\"ur Astronomie, K\"onigstuhl 17, D-69117, Heidelberg, Germany}
\affil{National Radio Astronomy Observatory, Pete V. Domenici Array Science Center, P.O. Box 0, Socorro, NM 87801, USA}

\author{Manuel Aravena}
\affil{N\'ucleo de Astronom\'ia de la Facultad de Ingenier\'ia y Ciencias, Universidad Diego Portales, Av. Ej\'ercito Libertador 441, Santiago, Chile}

\author[0000-0002-3952-8588]{Leindert Boogaard}
\affil{Leiden Observatory, Leiden University, P.O. Box 9513, NL-2300 RA Leiden, The Netherlands}

\author{Chris Carilli}
\affil{National Radio Astronomy Observatory, Pete V. Domenici Array Science Center, P.O. Box 0, Socorro, NM 87801, USA}

\author{Roberto Decarli}
\affil{INAF-Osservatorio di Astrofisica e Scienza dello Spazio, via Gobetti 93/3, I-40129, Bologna, Italy}

\author{Tanio D\'iaz-Santos}
\affil{Institute of Astrophysics, Foundation for Research and Technology--Hellas (FORTH), Heraklion, GR-70013, Greece}
\affil{N\'{u}cleo de Astronom\'ia, Facultad de Ingenier\'ia y Ciencias, Universidad Diego Portales, Av. Ej\'ercito Libertador 441, Santiago, Chile}
\affil{Chinese Academy of Sciences South America Center for Astronomy (CASSACA), National Astronomical Observatories, CAS, Beijing 100101, China}

\author{Yoshi Fudamoto}
\affil{Department of Astronomy, University of Geneva, Ch. des Maillettes 51, 1290 Versoix, Switzerland}

\author{Hanae Inami}
\affil{Hiroshima Astrophysical Science Center, Hiroshima University, 1-3-1 Kagamiyama, Higashi-Hiroshima, Hiroshima 739-8526, Japan}

\author{Rychard Bouwens}
\affil{Leiden Observatory, Leiden University, P.O. Box 9513, NL-2300 RA Leiden, The Netherlands}

\author[0000-0002-3583-780X]{Paulo C. Cortes}
\affil{Joint ALMA Observatory - ESO, Av. Alonso de C\'ordova, 3104, Santiago, Chile}
\affil{National Radio Astronomy Observatory, 520 Edgemont Rd, Charlottesville, VA, 22903, USA}

\author[0000-0003-2027-8221]{Pierre Cox}
\affil{Sorbonne Universit{\'e}, UPMC Universit{\'e} Paris 6 \& CNRS, UMR 7095, Institut d'Astrophysique de Paris, 98b boulevard Arago, 75014 Paris, France}

\author{Emmanuele Daddi}
\affil{Laboratoire AIM, CEA/DSM-CNRS-Universite Paris}

\author[0000-0003-3926-1411]{Jorge Gonz\'alez-L\'opez}
\affil{Las Campanas Observatory, Carnegie Institution of Washington, Casilla 601, La Serena, Chile}
\affil{N\'ucleo de Astronom\'ia de la Facultad de Ingenier\'ia y Ciencias, Universidad Diego Portales, Av. Ej\'ercito Libertador 441, Santiago, Chile}

\author{Ivo Labbe}
\affil{Centre for Astrophysics and Supercomputing, Swinburne University of Technology, Melbourne, VIC 3122, Australia}

\author{Gerg\"o Popping}
\affil{European Southern Observatory, Karl-Schwarzschild-Strasse 2, 85748, Garching, Germany}

\author{Dominik Riechers}
\affil{Department of Astronomy, Cornell University, Space Sciences Building, Ithaca, NY 14853, USA}
\affil{Max-Planck-Institut f\"ur Astronomie, K\"onigstuhl 17, D-69117 Heidelberg, Germany}

\author{Mauro Stefanon}
\affil{Leiden Observatory, Leiden University, P.O. Box 9513, NL-2300 RA Leiden, The Netherlands}

\author{Paul Van der Werf}
\affil{Leiden Observatory, Leiden University, P.O. Box 9513, NL-2300 RA Leiden, The Netherlands}

\author[0000-0003-4678-3939]{Axel Weiss}
\affil{Max-Planck-Institut f\"ur Radioastronomie, Auf dem H\"ugel 71, 53121 Bonn, Germany}

\begin{abstract}   

The ALMA Spectroscopic Survey in the \emph{Hubble} Ultra Deep Field (ASPECS) Band 6 scan (212--272~GHz) covers potential [CII] emission in galaxies at $6\leq z \leq8$ throughout a 2.9~arcmin$^2$ area. By selecting on known Lyman-$\alpha$ emitters (LAEs) and photometric dropout galaxies in the field, we perform targeted searches down to a 5$\sigma$ [CII] luminosity depth $L_{\ciimath}\sim2.0\times10^8$~L$_{\odot}$, corresponding roughly to star formation rates (SFRs) of $10$--20~M$_{\odot}$~yr$^{-1}$ when applying a locally calibrated conversion for star-forming galaxies, yielding zero detections. While the majority of galaxies in this sample are characterized by lower SFRs, the resulting upper limits on [CII] luminosity in these sources are consistent with the current literature sample of targeted ALMA observations of $z=6$--7 LAEs and Lyman-break galaxies (LBGs), as well as the locally calibrated relations between $L_{\ciimath}$ and SFR---with the exception of a single [CII]-deficient, UV luminous LBG. We also perform a blind search for [CII]-bright galaxies that may have been missed by optical selections, resulting in an upper limit on the cumulative number density of [CII] sources with $L_{\ciimath}>2.0\times10^8$~L$_{\odot}$ ($5\sigma $) to be less than $1.8\times10^{-4}$~Mpc$^{-3}$ (90\% confidence level). At this luminosity depth and volume coverage, we present an observed evolution of the [CII] luminosity function from $z=6$--$8$ to $z\sim0$ by comparing the ASPECS measurement to literature results at lower redshift.
       
\end{abstract}

\section{Introduction} \label{sec:intro}

Characterizing the properties of the interstellar medium (ISM; dust and gas) of the first generations of galaxies is one of the prime goals in observational astrophysics: Given the likely role of early galaxies in cosmic reionization -- the last major phase transition of the Universe, which was completed by z$\sim$6 -- understanding their physical properties is of particular importance \citep[e.g.,][]{Dayal2018}. Evidence has emerged that the ISM conditions of pre--reionization galaxies were very different than in their descendants at later cosmic epochs. This includes strong rest-frame UV emission lines from ground-based spectra \citep[e.g., CIII\symbol{93};][]{Stark2015,Mainali2018} as well as extreme equivalent width optical lines measured via Spitzer colors \citep[\symbol{91}OIII\symbol{93}5007+H$\beta$;][]{Labbe2013,DeBarros2019}. All these measurements point to hard ionization fields, dominated by young, low-metallicity stars -- very different from galaxies at later times.

The sensitivity of ALMA now allows one to obtain more detailed insights into the chemical and physical properties of early galaxies at $z\geq 6$. In particular, measurements of the [CII]158$\mu$m line of the ISM provide unique constraints on the molecular gas properties and star-formation rates (SFRs) of galaxies \citep[e.g.,][]{diaz2013,HerreraCamus2015}. [CII] is often the dominant cooling line of the ISM, coming primarily from photo-dissociation regions and the cold neutral medium of molecular clouds \citep[e.g.,][]{Vallini2013}. As such, the [CII] line probes the gas, from which stars are formed in normal galaxies \citep[][]{CarilliWalter2013,DeLooze2014,Zanella2018}. 

[CII] lies in a favorable frequency window for $6\leq z \leq 8$ galaxies (ALMA band~6). Even though a very large number of $z>6$ galaxies have now been identified from deep HST imaging \citep[e.g.,][]{Bouwens2015,Finkelstein2015}, only a small number of the brightest galaxies have been spectroscopically confirmed via their Ly$\alpha$ emission lines \citep[e.g.,][]{Oesch2015,Zitrin2015}, due to a higher IGM opacity at $z>6$ in the neutral era of the universe \citep[e.g.,][]{Schenker2012,Treu2013,Pentericci2014}. [CII] detections with ALMA therefore promised to be an efficient new avenue to spectroscopically confirm high-redshift galaxies with missing Ly$\alpha$ emission.

[CII] has now been detected in several non-quasar host galaxies at $z>6$ \citep[see e.g.,][]{Maiolino2015,Willott2015,Hashimoto2019,Bakx2020}. However, its luminosity was often not as high as expected compared to the local relation between SFR and $L_{\ciimath}$ \citep[][]{DeLooze2014,HerreraCamus2015}. While relatively luminous [CII] emission is still found at z$\sim$4.5 - 5.5 \citep[e.g.][]{Capak2015,Schaerer2020}, evidence is building for a deficit in $L_{\ciimath}$ and an evolution of the SFR--$L_{\ciimath}$ relationship at $z>6$ in the epoch of reionization \citep[see e.g.,][]{Pentericci2016,Matthee2019, Laporte2019,Harikane2020}. Theoretically, this can be well explained with lower metallicities expected in early galaxies \citep[][]{Vallini2013,Lagache2018,Popping2019}, and high surface densities of star formation in starbursting galaxies \citep{Ferrara2019_ciisfr}. Recent observations and re-analyses of earlier ALMA data have now led to a different possible scenario: a significantly larger scatter in $L_{\ciimath}$ at high-redshift compared to the local SFR--$L_{\ciimath}$ relation, instead of a [CII] deficit \citep[see e.g.,][]{Carniani2018,Matthee2019}. Additionally, surface brightness dimming could affect the detectability of [CII] emission \citep[][]{Carniani2020}.
However, the current datasets are still limited in size and, furthermore, most of the early galaxies that have been observed with ALMA were selected as Ly$\alpha$ emitters (such that they had a previously known redshift), which can lead to a bias toward young, metal-poor, dust-free systems \citep[see e.g.,][]{Smit2018}.

The ALMA large program ASPECS provides the first full frequency scan in band 6 of the Hubble Ultradeep Field \citep[HUDF;][]{Decarli2019_3mm,Aravena2019_3mm,GL2019_3mm}. These observations enable the unbiased search for emission lines, both molecular (CO) and atomic ([CI]) (\citet{Decarli2020_1mm} and \citet{Boogaard2020_1mm}) and [CII] (this paper). The HUDF was chosen as it represents the deepest dataset available across all wavelengths. By design, this field does not include very massive and highly star-forming systems (such as submillimeter galaxies or quasars), but traces the field galaxy population that is most representative at each cosmic epoch ($L^{\star}$ and sub-$L^{\star}$ galaxies). The HUDF has been particularly important in the discoveries of the most distant galaxies known, from early studies of the $z>6$ galaxy population \citep[e.g.,][]{Bouwens2010,McLure2010,Oesch2010} to consecutively higher redshifts \citep[now extending to z$\sim$10--12; e.g.,][]{Ellis2013,Oesch2013}. The frequency setup of the ASPECS band~6 scan covers the redshifted [CII] emission line from $6<z<8$, ideally matched to some of the most distant galaxies known in the HUDF. In the pilot observations of ASPECS, a number of potential [CII] line candidates were reported \citep{Aravena2016_cii}. However, as discussed in detail below, none of these candidates could be confirmed (at 5$\sigma$) in the deeper and more uniform observations obtained through the ASPECS large program. This is in line with other recent [CII] detections and upper limits that have been reported in other sources in the meantime.

In this paper, we exploit these deeper data from the ASPECS large program over the HUDF to constrain the [CII] emission from galaxies at $z=6-8$. In particular, we search for [CII] emission of previously identified LAEs and LBGs in this field, also exploiting recent, very deep MUSE spectra \citep{Inami2017}, and we perform an additional blind search. Doing this, we constrain the SFR-$L_{\ciimath}$ relationship as well as the [CII] luminosity function in the epoch of reionization.
This paper is structured as follows: In Sec.~\ref{sec:obs} we describe the observational data that was used. Sec.~\ref{sec:results} presents the results of the [CII] emission line searches, before we discuss their implications in Sec.~\ref{sec:discussion}. We finish with conclusions in Sec.~\ref{sec:conclusions}.

Throughout this paper we use a concordance cosmology with $\Omega_M=0.3,~ \Omega_\Lambda=0.7, ~h=0.7$. Magnitudes are presented in the AB system \citep{Oke83}, and we use a Chabrier initial mass function \citep[IMF][]{Chabrier2003}. 

\section{Observations} \label{sec:obs}
\subsection{ASPECS LP Band 6 data}

ASPECS Band 6 data is presented in detail in, e.g., \citet{Decarli2020_1mm} and \citet{GL2020_1mm}. In brief summary, observations were conducted from March--April 2017 and May--July 2018, surveying a 4.2~arcmin$^2$ area (at 10\% mosaic primary beam response) in the UDF with 85 ALMA pointings and a total observing time of $\sim80$~hours, including overheads. During observing, the 12-meter array was in either compact configuration C40-1 or C40-2 to ensure galaxies were mostly or entirely spatially unresolved. The survey scanned the full bandwidth of ALMA B6 in 8 non-overlapping frequency setups, providing continuous wavelength coverage from observed frequencies $\nu_{obs}=212$--272~GHz. At these frequencies, redshifted [CII] emission can, in principle, be observed from redshifts $z = 5.99$--7.97.

\begin{figure}
    \centering
    \includegraphics[width=0.45\textwidth]{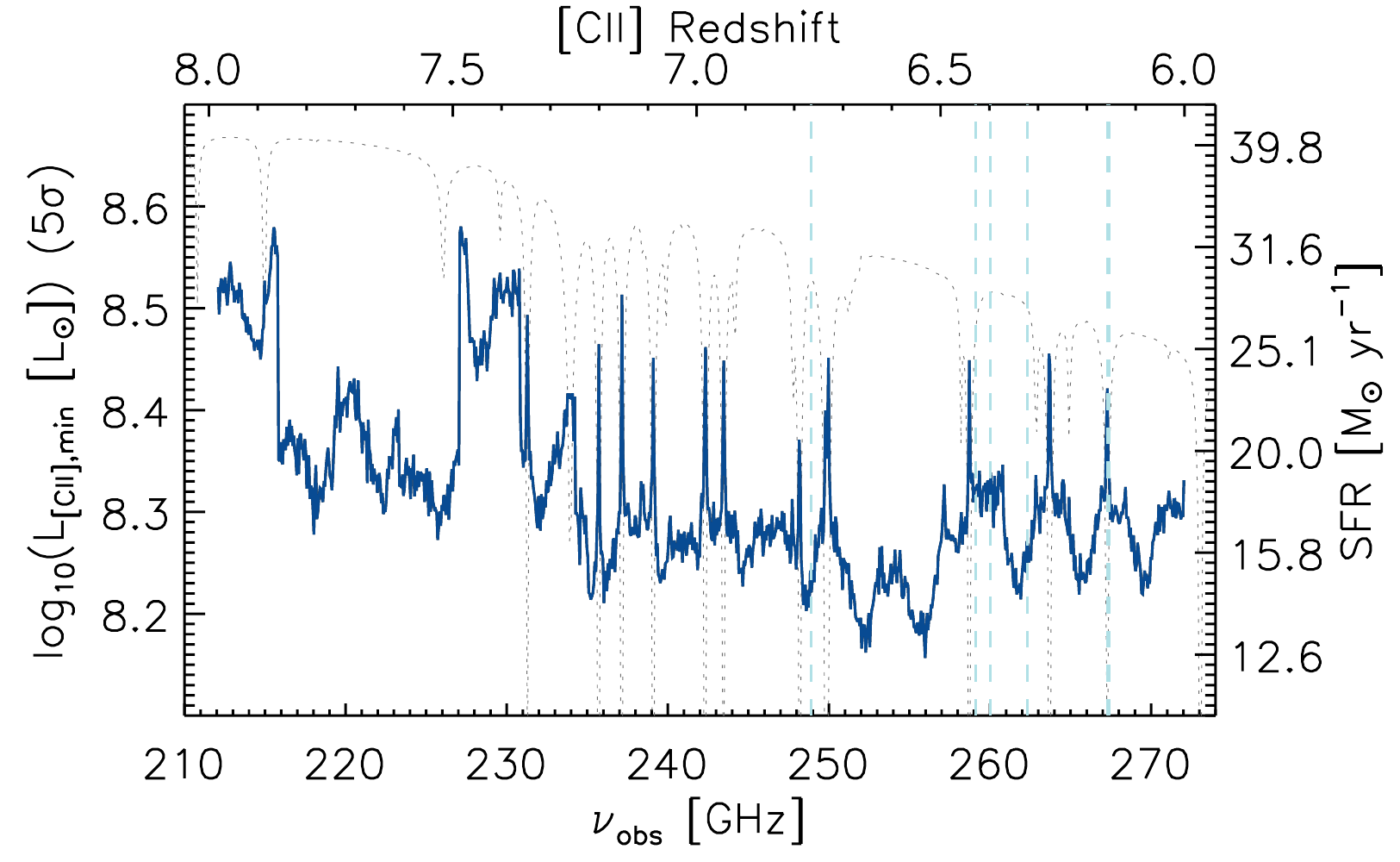}
    \caption{[CII] luminosity depth ($5\sigma$), assuming 200~km~s$^{-1}$ line width, across the survey bandwidth. The corresponding limit in SFR (right ordinate) is derived using a locally calibrated $L_{\ciimath}$-SFR relation for star-forming galaxies \citep{DeLooze2014}. Dashed (light blue) vertical lines indicate the expected $\nu_{obs}$ for the [CII] line based on the Ly$\alpha$ redshift for the MUSE LAEs in the sample defined in Section \ref{sec:ancillary_data}: MUSE 852, 6332, 6312, 6524, 802, 560, in order of decreasing redshift. (Note that the vertical lines for MUSE 802 and 560 at the far right are nearly overlapping due to the proximity in redshift.) For illustrative purposes only, we have also plotted the atmospheric transmission (assuming PWV = 1 mm; dark gray, dotted curve) to show the effect of the atmospheric absorption features on the survey depth.}
    \label{fig:ciilumlim}
\end{figure}

Throughout this work, we make use of two data products resulting from the ASPECS Band 6 survey. 
For extracting spectra, we use the naturally-weighted raw, or ``dirty," image cube, after applying a primary beam (PB) correction. Continuum from bright 1~mm sources has been subtracted from this cube, as described in Gonzalez-Lopez et al. 2020. We mask out data below the half power beam width (HPBW), where the mosaic primary beam response is less than 50\%, working only within the central 2.9~arcmin$^2$ in the survey footprint. The synthesized beam in the image cube is $\sim$1\farcs6 $\times$ 1\farcs1 at bandcenter $\nu_{obs,cen} = 242$~GHz, and the pixel scale is 0\farcs2 per pixel. We have rebinned frequency channels by a factor of 8, so that the spectral resolution is 62.5~MHz ($\sim77$~km~s$^{-1}$ at $\nu_{obs,cen}$). The resulting mean RMS is 0.30~mJy~beam$^{-1}$ per channel. To convert this flux density to an equivalent line luminosity, we assume spatially unresolved emission and adopt a fiducial line velocity width $v_{\mathrm{FWHM}} = 200$~km~s$^{-1}$---representative of observed line widths (FWHM) for [CII] emission in $z\sim6$--7 LAEs \citep[cf. Table C.1 in][]{Matthee2019}---to calculate flux (in units of Jy~km~s$^{-1}$), then divide the measured RMS by a scale factor $\sqrt{v_{\mathrm{FWHM}}/v_{chn}}$ to account for the number of spectral channels with velocity resolution $v_{chn}$ that spans $v_{\mathrm{FWHM}}$.\footnote{$\sqrt{v_{\mathrm{FWHM}}/v_{chn}} = 1.6$ at band center (242~GHz).} The $5\sigma$ line luminosity depth for the [CII] line as a function of observed frequency is shown in Figure~\ref{fig:ciilumlim}, where the different features in the sensitivity arise due to a combination of integration time at different frequencies and atmospheric transmission (see also \citet{Decarli2020_1mm}). For reference, imaging of the ASPECS Pilot data cube resulted in an average RMS level of 0.42~mJy~per beam over the same channel width \citep{Aravena2016_cii}. Combining both data sets results in a marginal increase---by only a factor $\sim1.14$---in sensitivity over the relevant area overlapping between the LP and Pilot survey fields. We thus decided to proceed with the independent datasets.

Additionally, we use the 1.2~mm continuum map (without PB correction) with a mean RMS of 9.3~$\mu$Jy beam$^{-1}$ presented in \citet{GL2020_1mm} to search for continuum emission in targeted optically-selected galaxies (described in the following Section~\ref{sec:ancillary_data}) and in positions returned by the blind search algorithm described in Section~\ref{sec:blind}. 

\subsection{Ancillary Datasets} \label{sec:ancillary_data}

To enable a targeted [CII] search in the ASPECS Band 6 data cube, we exploit existing galaxy catalogs in UDF with reliable photometric or spectroscopic redshifts.
\paragraph{Lyman-$\alpha$ emitters (LAEs)} The MUSE \emph{Hubble} Ultra Deep Field survey \citep{Bacon2017} provides accurate spectroscopic redshifts for $\sim1500$ galaxies in a 9~arcmin$^2$ field encompassing the full ASPECS footprint. With its wide instantaneous bandwidth, spanning 4650--9300~\AA, the MUSE IFU spectrometer provides continuous coverage of the Lyman-$\alpha$ (Ly$\alpha$) emission line from $z=2.8$--6.7; we refer the interested reader to, e.g., \citet{Boogaard2019_3mm} for more information on observational details of the MUSE survey and complementarities with ASPECS. We consider sources located within the ASPECS HPBW and which have secure spectroscopic redshifts within the ASPECS [CII] redshift coverage. These criteria yield 6 LAEs in our sample at $z=6.1$--6.6, drawn from the spectroscopic redshift catalog presented in \citet{Inami2017}. The LAEs targeted here are characterized by Ly$\alpha$ fluxes $F_{\mathrm{Ly}\alpha} = 1.5$--$11.2\times 10^{-18}$~erg~s$^{-1}$~cm$^{-2}$, corresponding to line luminosities $L_{\mathrm{Ly}\alpha}= 0.66$--$1.5\times10^{42}$~erg~s$^{-1}$, where $10^{42}$~erg~s$^{-1}$ is approximately $0.2L^{*}_{\mathrm{Ly}\alpha}$ \citep{Drake2017} at these redshifts. Their rest-frame Ly$\alpha$ equivalent widths (EWs) span a wide range, between $\sim7$--140~\AA~(private communication, T. Hashimoto). 

\paragraph{Lyman-break galaxies (LBGs)} We use the most comprehensive sample of $z>6$ Lyman-break selected galaxies in the XDF and GOODS-S fields that overlap with the ASPECS footprint, originally presented in \citep{Bouwens2015}. Only sources within the ASPECS HPBW and with photometric redshifts where the redshift probability distribution functions, $p(z)$, have $>68\%$ confidence to lie within the ASPECS [CII] frequency coverage are included in our sample.  Regarding the latter criterion, we require the 1$\sigma$ lower and upper limits on the peak redshift, $z_{peak}$, determined from the $p(z)$, to be $> 6$ or $<8$, respectively. In total, there are 45 LBGs that satisfy these criteria at $6.1 \leq z_{peak} \leq 7.6$ with \emph{HST} F160W band, or $H$ band, magnitudes ranging from the 6 brightest LBGs in ASPECS with $H_{160} < 27.5$ mag to the faintest at $H_{160}=30.9$ mag, which corresponds roughly to the 5$\sigma$ depth in XDF. The uncertainty on $z_{peak}$ ranges from $\Delta z_{peak}=0.05$--0.52 (1$\sigma$), with a median value $\sim0.25$ across the sample.\footnote{For reference, the minimum and maximum $\Delta z_{peak}$ correspond to uncertainties of $\pm3.1$ and $\pm8.5$~GHz, respectively, in the observed frame, centered at the the expected frequency for [CII] at $z_{peak}$.} 

\subsubsection{Properties derived from SED fitting}

For all LAEs in the sample, we measured the HST photometry in the four WFC3/IR filters (F105W, F125W, F140W, and F160W) from the XDF postage stamps \citep{Illingworth2013} in 0\farcs4 radius apertures. For all LBGs, HST photometry is available from the original selection paper. Additionally, we measured IRAC photometry based on the latest reductions of all the Spitzer/IRAC imaging available in the GOODS field as part of the GREATS survey (Stefanon et al. 2020, in preparation).

These measurements were used to derive star-formation rates as well as photometric redshifts (for the LBGs only) based on spectral energy distribution (SED) fits using the codes EAzY \citep{Brammer2008} and FAST++ (Schreiber et al., in prep.\footnote{A rewrite of the original FAST IDL code \citep{Kriek09} in C++ available at \url{https://github.com/cschreib/fastpp}}), respectively. (For the LAEs, the redshift was kept fixed at the Ly$\alpha$ redshift.) We adopt \citet{BC03} models with metallicities of 0.2~$Z_{\odot}$, constant star-formation histories, and a \citet{Calzetti1997} dust law to derive SED-based star-formation rates for all sources. This approach is very similar to what has been used in \citet{Bouwens2020}.

\section{Results} 
\label{sec:results}

\subsection{[CII] search in optically/near-IR selected galaxies} \label{sec:optical_nir}

\subsubsection{Lyman-$\alpha$ emitters} \label{sec:lae}

We obtain Band 6 spectra for the LAEs using a single-pixel extraction (equivalent to an extraction over the area of the synthesized beam) at the source position determined from \emph{HST} photometry \citep{Inami2017}; at $6 \lesssim z \lesssim 8$, sources are generally expected to be unresolved by our synthesized beam size of $1\farcs6\times 1\farcs1$ (= 8.4~kpc $\times$ 5.7~kpc at $z=7.0$). Due to known astrometric offsets between ALMA and \emph{HST} data, we measure source coordinates using the Hubble Legacy Field (HLF) reduction of the GOODS-S field that has been shifted to match \emph{Gaia} data \citep{Whitaker2019}. Comparing positions measured using the original \emph{HST} coordinates and the Gaia-matched coordinates for our sources, we find median offsets $\Delta \mathrm{(RA)}_{HST} = 0\farcs14$ and $\Delta\mathrm{(Dec)}_{HST} =-0\farcs24$, consistent with findings from previous ALMA data over this field \citep{Dunlop2017,Franco2018}. These shifts are smaller than or comparable to the $0\farcs2$ pixel size in the ASPECS 1mm image cube. 

\begin{table*}
    \centering
    \caption{Source properties for MUSE LAEs in ASPECS LP}
    \begin{tabular}{c c c c c c c}
    \hline \hline 
    MUSE ID & $z$ & RA & Dec & SFR$_{SED}$   &$L_{\mathrm{Ly}\alpha}$ & $L_{\ciimath}$ \\
            &    & [deg]      &  [deg]           &  [M$_{\odot}$~yr$^{-1}$]  & [$10^{42}$ erg s$^{-1}$]   & [$10^8$ L$_{\odot}$] \\ 
    (1)     &     (2) &    (3)     &  (4)         & (5)                       & (6)                   &   (7)          \\ 
    \hline
    852   & 6.636  & 53.169048   & $-27.778835$   &   1.20$^{+0.09}_{-0.18}$   & 1.29  & $<2.24$   \\ 
    6312  & 6.310  & 53.166118   & $-27.772048$   &   4.79$^{+3.92}_{-2.16}$   & 5.03 & $<2.47$   \\ 
    802   & 6.110  &  53.168540  & $-27.775677$    &   0.18$^{+0.04}_{-0.03}$   & 1.45 & $<2.15$   \\ 
    6332  & 6.335  & 53.158161   & $-27.778554$   &   0.12$^{+0.42}_{-0.05}$   & 1.28 &$<2.25$   \\ 
    6524  & 6.245  & 53.158247   & $-27.767763$   &   0.16$^{+0.51}_{-0.10}$   & 0.66 &$<2.64$    \\ 
    560   & 6.107  & 53.159523    & $-27.771524$   &   13.49$^{+10.50}_{-5.90}$ & 1.48 & $<2.05$     \\ 
    \hline
    \multicolumn{7}{p{0.7\textwidth}}{---\emph{Notes:} (1) MUSE ID from \citet{Inami2017}. (2) Ly-$\alpha$ redshift, determined for the peak flux of the Ly-$\alpha$ profile (3), (4) RA and Dec determined from \emph{Hubble} XDF $Y$-band (F105W) image, including spatial offsets from Gaia-matched reduction the HLF GOODS-S images. (5) SED-based SFR estimate.  
    (6) Ly$\alpha$ luminosity in units of $10^{42}$~erg~s$^{-1}$ \citep{Inami2017}. (7) Upper limit (5$\sigma$) on [CII] luminosity, in units of $10^8$~$L_{\odot}$, assuming $\mathrm{FWHM} = 200$~km~s$^{-1}$.
    }
    \end{tabular}
    \label{tab:lae_props}
\end{table*}

\begin{figure*}
    \centering
    \begin{tabular}{c c}
    \includegraphics[width=0.45\textwidth]{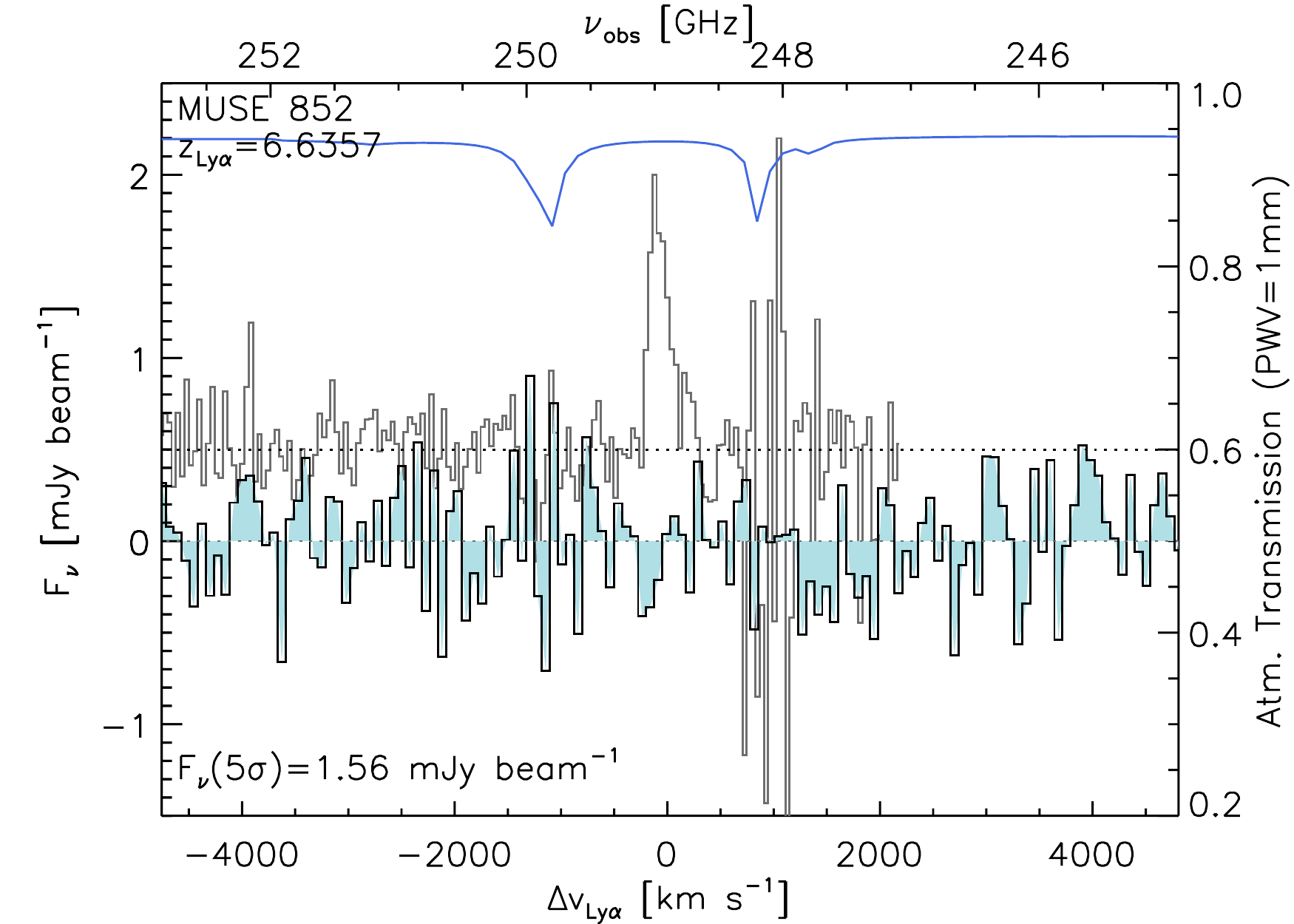} &
    \includegraphics[width=0.45\textwidth]{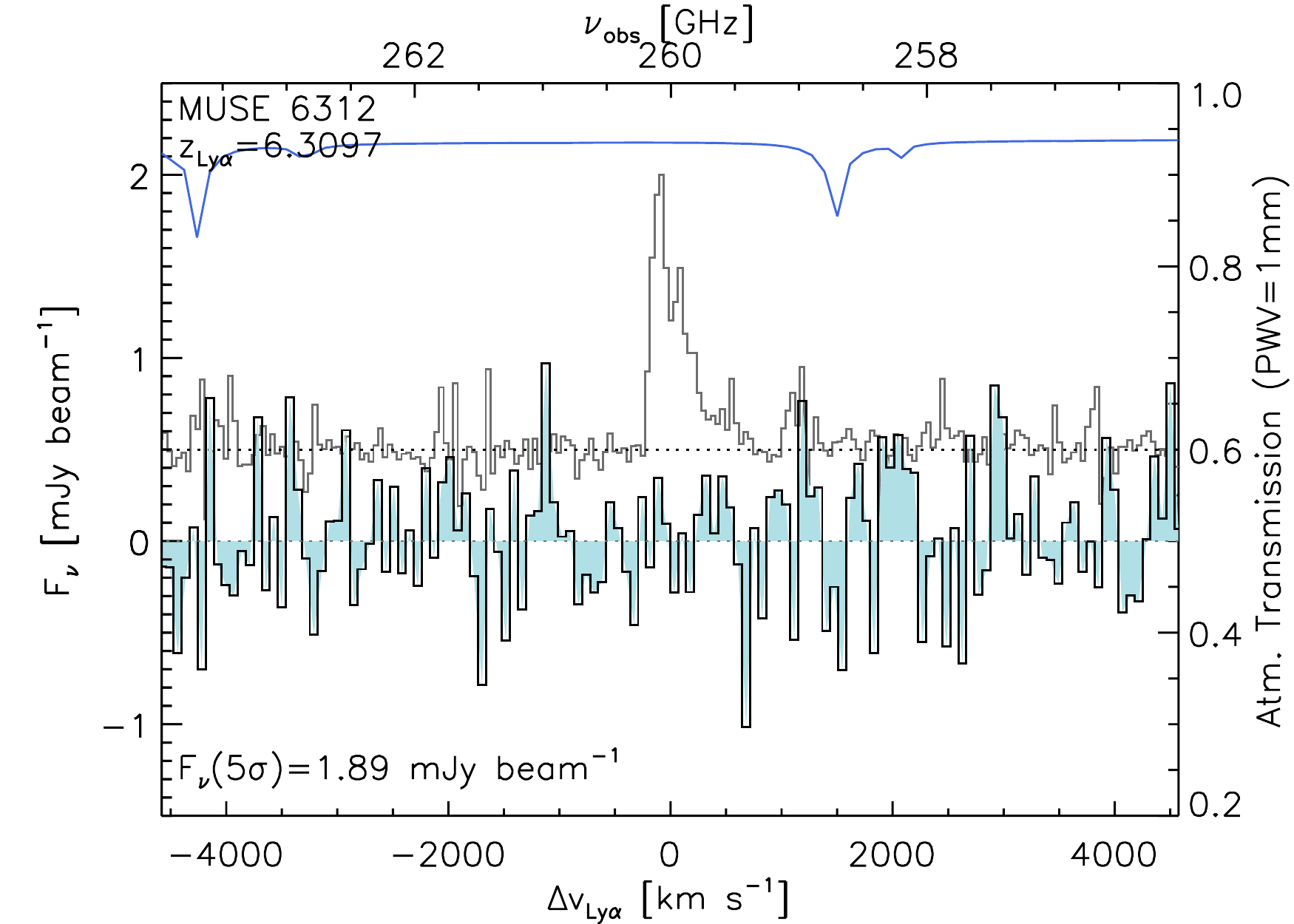} \\
    \includegraphics[width=0.45\textwidth]{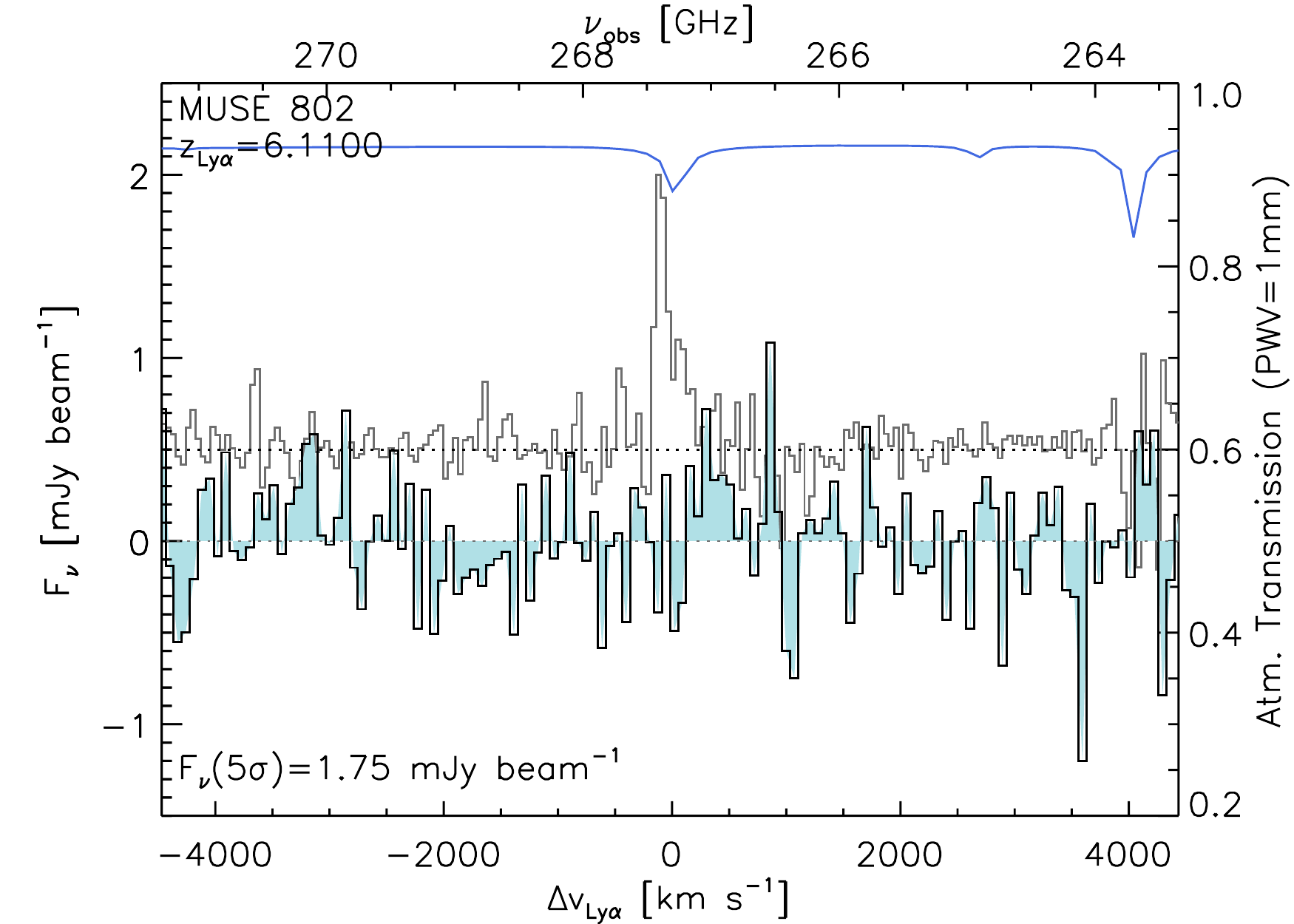} &
    \includegraphics[width=0.45\textwidth]{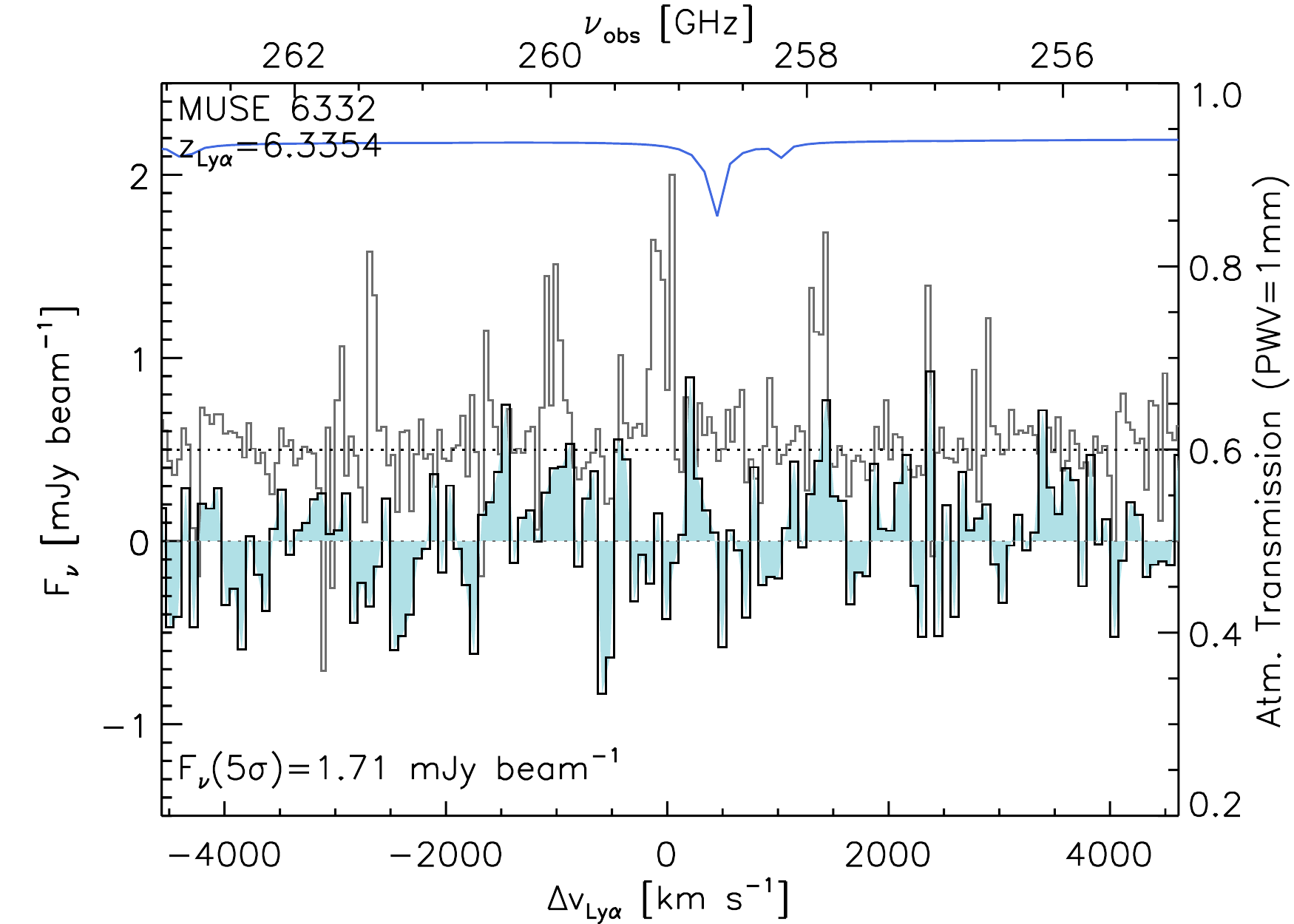} \\
    \includegraphics[width=0.45\textwidth]{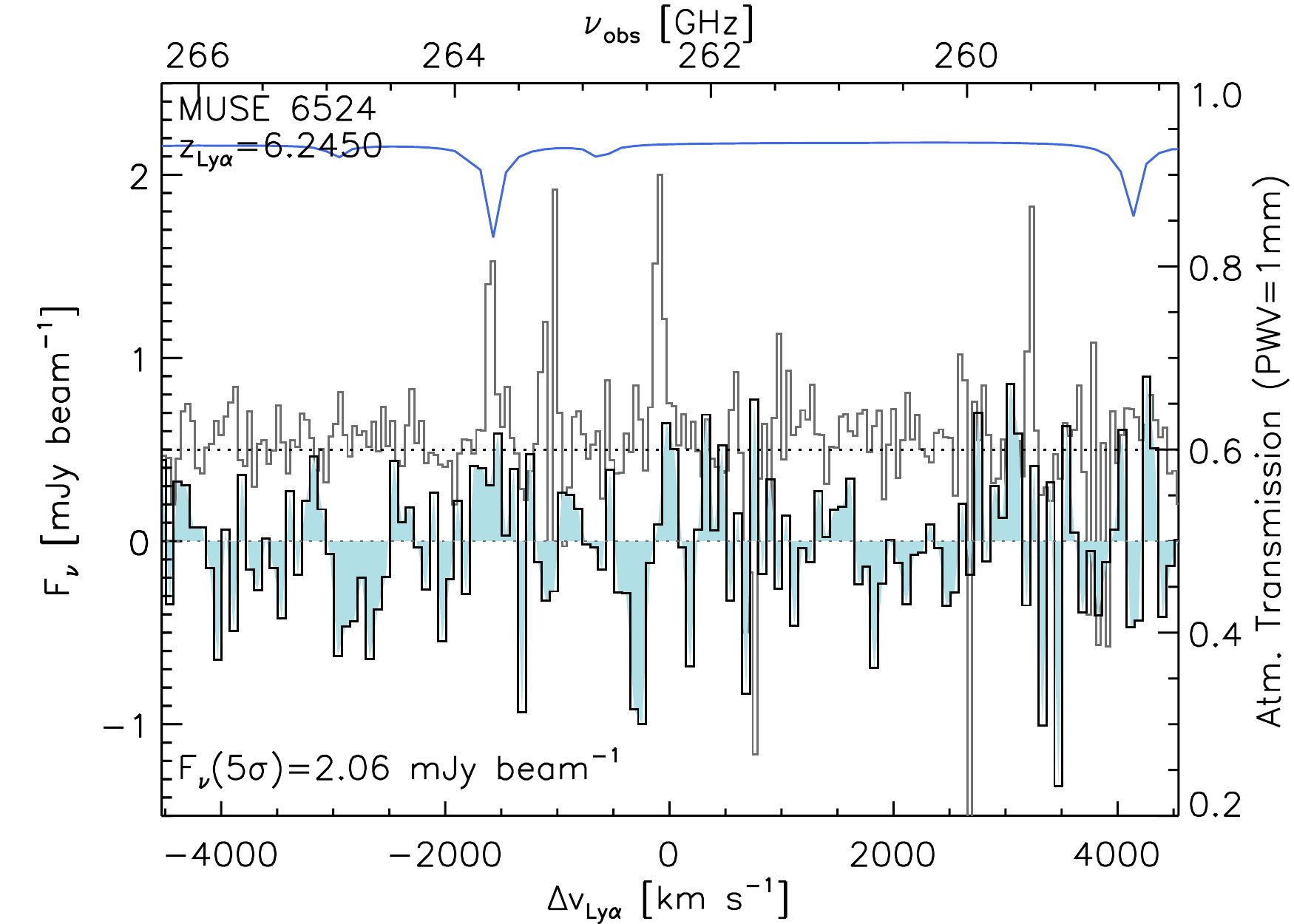} &
    \includegraphics[width=0.45\textwidth]{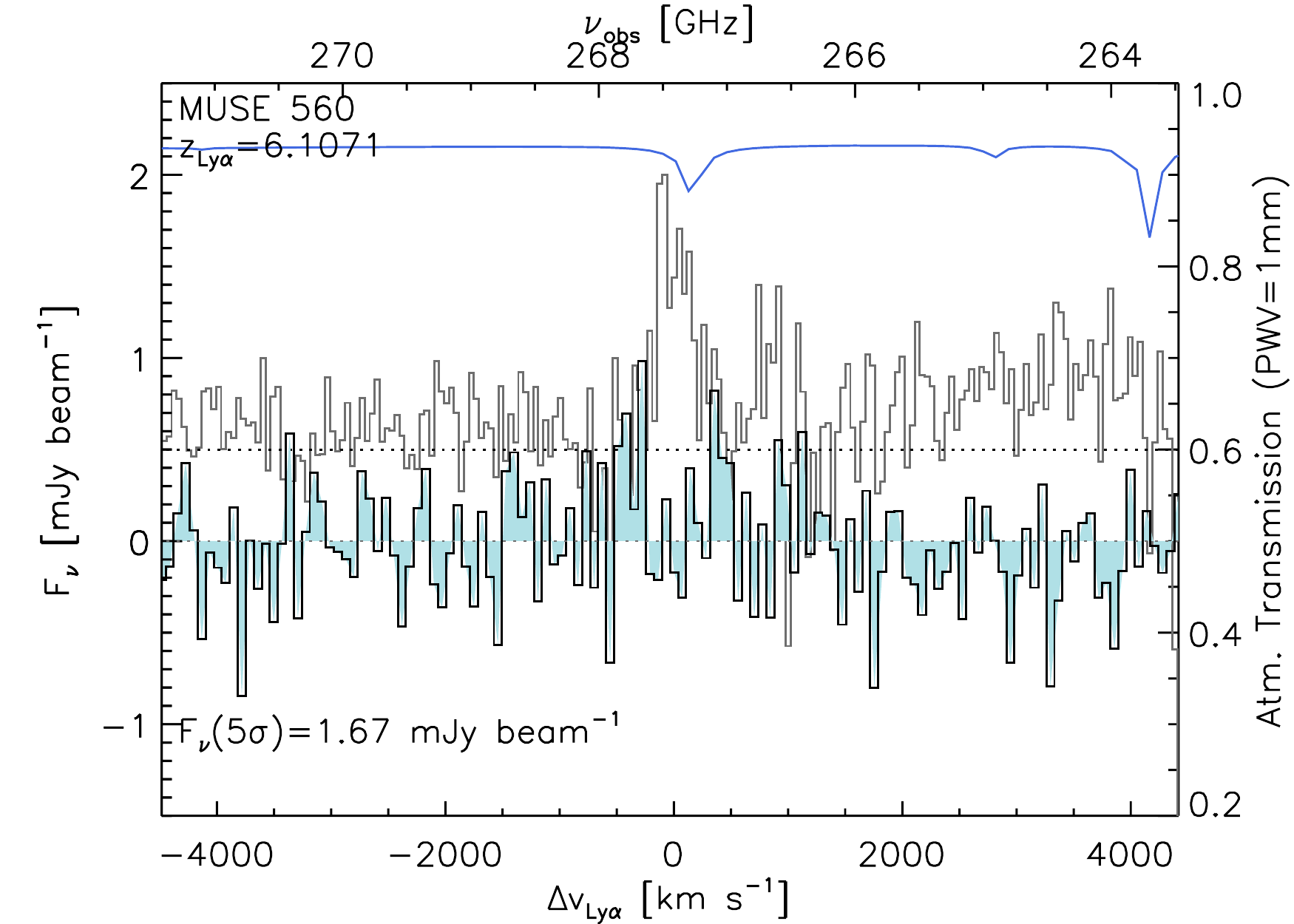} \\
    \end{tabular}
    \caption{Band 6 spectra ($\Delta\nu_{chn}=62.5$~MHz) extracted at positions of known LAEs with secure redshifts (CONFID $\ge 2$) from \citet{Inami2017} within the ASPECS LP spectral and spatial survey coverage. Lyman-$\alpha$ spectrum for each source is overplotted in gray, with arbitrary flux density scaling in each panel. For reference, we also show atmospheric transmission assuming PWV = 1.0~mm (solid blue curve).}
    \label{fig:lae_spectra}
\end{figure*}

Resulting spectra are shown in Figure~\ref{fig:lae_spectra} as a function of the offset in velocity units from the Lyman-alpha redshift $z_{\mathrm{Ly}\alpha}$,  $\Delta v_{\mathrm{Ly}\alpha} = c(z_{\ciimath}-z_{\mathrm{Ly}\alpha})/(1+z_{\ciimath})$, where $z_{\ciimath}$ corresponds to the expected [CII] redshift at the observed frequency. The [CII] line is not detected in any of the six LAEs in our sample. In order to place upper limits on [CII] luminosities for these sources, we first measure the RMS in flux density across 8~GHz of bandwidth in the spectrum, centered at the expected frequency for redshifted [CII] emission based on the Ly$\alpha$ redshift. As in Section~\ref{sec:obs}, we use the fiducial $v_{\mathrm{FWHM}}= 200$~km~s$^{-1}$ when converting the RMS in flux density to a corresponding limit in line luminosity. Source properties, including our derived limits on [CII] luminosities $L_{\ciimath}$, are summarized in Table~\ref{tab:lae_props}.

\begin{figure}
    \centering
    \includegraphics[width=0.45\textwidth]{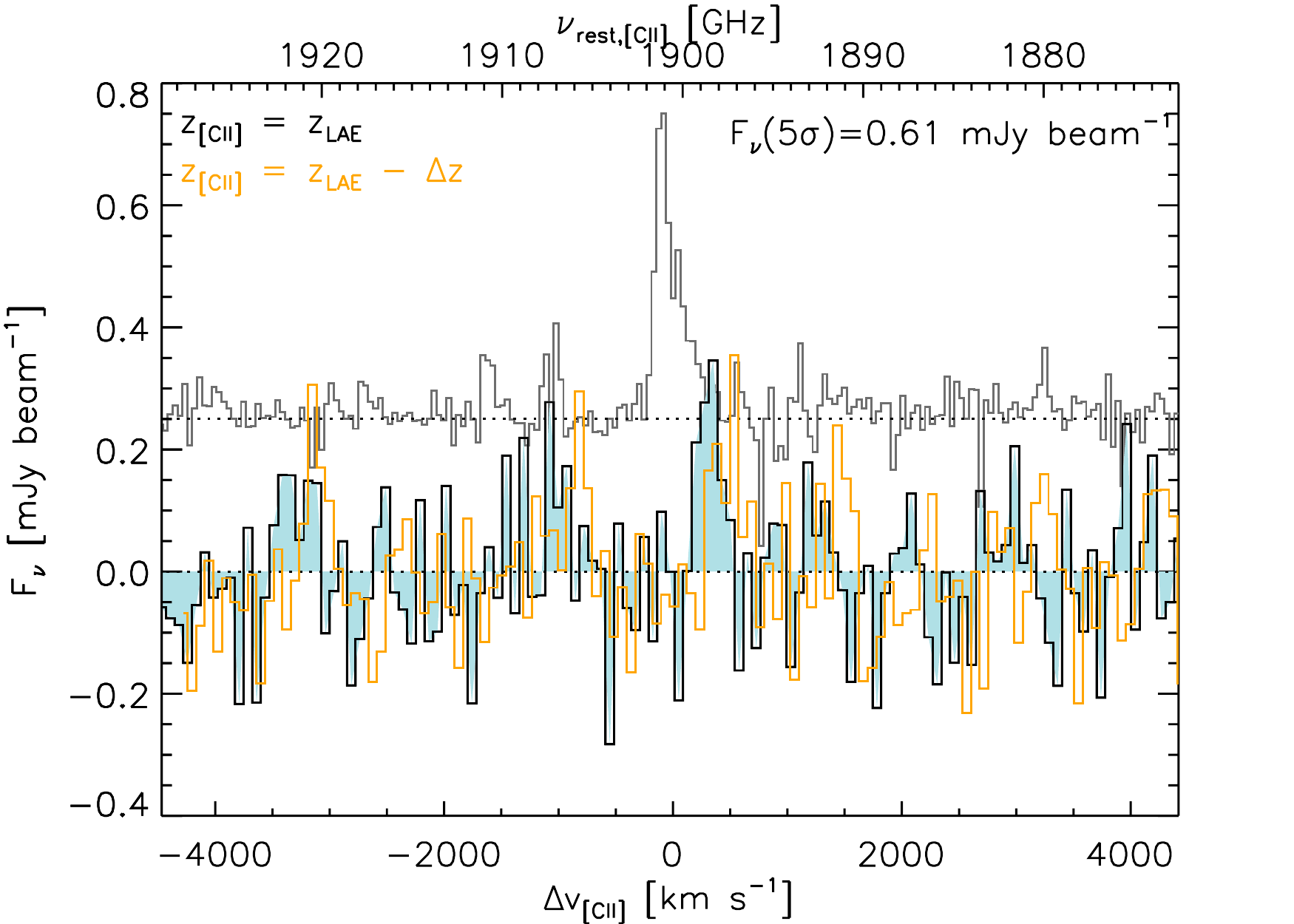}
    \caption{Stacked B6 spectra for the 6 MUSE LAEs in ASPECS, assuming that the [CII] redshift matches exactly the Ly$\alpha$ redshift (blue filled curve) or an empirically derived systemic redshift per \citep{Verhamme2018} (orange curve). The stacked Ly$\alpha$ spectrum is also shown for comparison (gray curve), with arbitrary flux density normalization. The 5$\sigma$ rms $F_{\nu}(5\sigma) = 0.61$~mJy~beam$^{-1}$ reported here refers to the B6 stacked spectrum computed without a velocity offset for the [CII] line.}
    \label{fig:lae_spectra_stack}
\end{figure}

Stacking the six spectra at $\Delta v_{\ciimath} = 0$~km~s$^{-1}$, we obtain an average, unweighted MUSE LAE B6 spectrum to search for faint emission. The stack was performed by first blueshifting the B6 spectra to their rest-frame frequencies, adopting either $z_{\ciimath}=z_{\mathrm{Ly}\alpha}$ or $z_{\ciimath}=z_{\mathrm{Ly}\alpha} - \Delta z_{\mathrm{Ly}\alpha}$ for the conversion. Here, $\Delta z_{\mathrm{Ly}\alpha}$ represents an offset between the Ly$\alpha$ redshift and the systemic redshift of the galaxy, $z_{sys}$, traced by [CII]. To determine $\Delta z_{\mathrm{Ly}\alpha}$, we follow the approach adopted in ASPECS LP Band 3 CO line stacking \citep{Inami2020}, using the empirical correlation between the FWHM of the Ly$\alpha$ line profile (measured in \citet{Inami2017}) and the redward velocity shift of Ly$\alpha$ with respect to $z_{sys}$, presented in \citet{Verhamme2018}. For MUSE 852, 6312, 802, 6332, 6524, and 560, the respective Ly$\alpha$ $\mathrm{FWHM}=9.0$, 9.4, 4.8, 4.7, 4.0, and 7.0~\AA. The resulting [CII] velocity offsets per \citet{Verhamme2018} (their Equation~2) are 230, 250, 120, 110, 89, and 180~km~s$^{-1}$ blueward of the Ly$\alpha$ line peak, or, equivalently, $\Delta z_{\mathrm{Ly}\alpha} = 0.0058$, 0.0061, 0.0028, 0.0026, 0.0021, and 0.0044. After converting the spectra to the rest-frame, each spectra was linearly resampled onto a reference frequency grid spanning the same velocity range ($\pm4,000$~km~s$^{-1}$, or 56.33~GHz in the rest-frame frequency axis) as for the individual spectra, with velocity resolution chosen to reflect the coarsest resolution of $75.30$~km~s$^{-1}$ (or 0.4773~GHz in the rest-frame) occuring for the highest redshift source, MUSE 852. The simple arithmetic average\footnote{An inverse-variance weighted average, where variances were determined on a channel by channel basis for each MUSE LAE at their spatial positions in the data cube, was also attempted. The resulting stacked spectrum was similar to the unweighted average, which we have adopted for simplicity.} of the resampled spectra was used to derive the final stacked spectrum for each choice of $z_{\ciimath}$ (i.e., with and without the velocity shift), presented in Figure~\ref{fig:lae_spectra_stack}. At an average redshift $\langle z_{\ciimath} \rangle=6.29$, the $5\sigma$ rms, 0.61~mJy~beam$^{-1}$, across the full $4,000$~km~s$^{-1}$ bandwidth in the stacked spectrum for $z_{\ciimath}=z_{\mathrm{Ly}\alpha}$ corresponds to an upper limit $L_{\ciimath} < 7.6\times10^7$~L$_{\odot}$.

\subsubsection{Lyman-break galaxies} \label{sec:lbg}

We have extracted single-pixel spectra in the ASPECS Band 6 data for the 45 Lyman-break selected sources described in Section~\ref{sec:ancillary_data}. For consistency with the noise estimation on the LAE spectra, we extract spectra in a local bandwidth of 8~GHz, centered at the expected observed frequency for [CII] based on the peak redshift, $z_{peak}$, from the $p(z)$ for a given LBG. In cases where the 1-$\sigma$ uncertainty on $z_{peak}$ is greater than 8~GHz, we use the upper and lower 1-$\sigma$ limits to determine the appropriate frequency range. 

No detections are reported. Upper limits on [CII] line luminosity are presented in Table~\ref{tab:lbg_props} for a subset of the LBG sample containing 5 of the brightest objects with derived SFR $\gtrsim10$, roughly corresponding the average 5$\sigma$ depth in SFR for the ASPECS [CII] survey (cf. Figure~\ref{fig:ciilumlim}).

\begin{table*}
    \centering
    \caption{Source properties for LBGs in ASPECS LP with SFR$_{SED}>10$~M$_{\odot}$~yr$^{-1}$}
    \begin{tabular}{c c c c c c}
    \hline \hline 
   ID & $z$ & RA & Dec & SFR$_{SED}$ & $L_{\ciimath}$ \\
       &    & [deg] &  [deg]  &   [M$_{\odot}$~yr$^{-1}$]  & [$10^8$~L$_{\odot}$]    \\ 
   (1) & (2)&(3)  &    (4)   &    (5)                 &      (6)         \\ 
    \hline
    XDFI-2374646327 & 6.48$\pm0.07$ & 53.156096 & $-27.775775$ &48$^{+8}_{-9}$ & $< 2.0$ \\
    XDFZ-2425646566 &  6.83$\pm0.06$  & 53.177333 & $-27.782389$ & 26$^{+6}_{-4}$ & $< 2.6$ \\
    XDFY-2395371744 & 7.58$\pm0.10$ & 53.164733 & $-27.788178$ &  20$^{+7}_{-3}$ & $<2.4$ \\
    XDFY-2388047071 & 7.54$\pm0.10$ & 53.161683 & $-27.785322$ &  19$^{+3}_{-7}$ & $<2.1$ \\
    GSDI-2382846172 & 6.08$\pm0.08$  & 53.159504 & $-27.771450$ & 12$^{+5}_{-4}$ & $<2.2$ \\
    \hline
    \multicolumn{6}{p{0.75\textwidth}}{---\emph{Notes:} (1), (2) ID and photometric redshift from \citet{Bouwens2015} (3) RA (4) Dec (5) SED-based SFR estimate from FAST++.  (6) Upper limit (5$\sigma$) on [CII] luminosity, in units of $10^8$~$L_{\odot}$, assuming $\mathrm{FWHM} = 200$~km~s$^{-1}$.
    }
    \end{tabular}
    \label{tab:lbg_props}
\end{table*}

\subsection{Blind [CII] line search} \label{sec:blind}

A blind search was performed for all spectral lines---including [CII], as well as lower redshift CO and atomic carbon lines---within the ASPECS LP 1.2~mm survey frequency coverage\footnote{In the on-sky dimension, the search was restricted to the 2.6~arcmin$^2$ area where the mosaic primary beam response is greater than 50\%.} in \citet{Decarli2020_1mm}. We refer the reader interested in details regarding the blind search algorithm and assessment of reality of blindly detected lines to that work (and references, therein), providing here only a brief summary to cover key steps and highlight important changes implemented in the LP analysis since the ASPECS Pilot study \citep{Walter2016_survey, Aravena2016_cii}.

As described in \citet{Decarli2020_1mm}, the line search was conducted using the \textsf{findclumps} algorithm \citep{Walter2016_survey, Decarli2019_3mm, GL2019_3mm}. This algorithm applies a 1-dimensional (1D) top-hat convolution in the spectral dimension of the data cube and identifies both positive and negative peaks in the emission, assigning to each peak a signal-to-noise ratio (SNR) calculated by comparing the peak flux density to the rms noise in the map. The width of the top-hat filter is varied iteratively in each convolution to search for spectral features with different line widths. For emission line candidates with SNR~$ > 4$, the 1D spectrum is extracted to retrieve a Gaussian-fitted line flux. Line (equivalently, redshift) identification was performed by cross-matching line candidates with ancillary data (e.g., photometric and spectroscopic galaxy catalogs in HUDF exploited in Sections~\ref{sec:lae} and \ref{sec:lbg}, and/or spectroscopic redshift from the ASPECS 3~mm dataset); or, when a line candidate failed to match to a catalog position, the line redshift was assigned on a probabilistic basis, taking into account the cosmic volume sampled by each possible line and various empirical weights reflecting the expected relative strength of the emission line to CO(1-0). 

The fidelity of a line candidate is quantified using a probabilistic approach that compares the number of positive and negative fluctuations, $N_{pos}$ and $N_{neg}$, in the data for a given SNR and convolution kernel width $\sigma_{kernel}$: 
\begin{equation}
\mathrm{fidelity} = 1 - \frac{N_{neg}\left(\mathrm{SNR},\sigma_{kernel}\right)}{N_{pos} \left(\mathrm{SNR},\sigma_{kernel}\right),}
\label{eq:fidelity_LP}
\end{equation}
where the allowable fidelity range from 0 to 1 implies a 100\% to 0\% chance, resp., that there are negative line candidates in the data with the same SNR and $\sigma_{kernel}$.

This treatment of fidelity improves upon, e.g., the analysis in \citet{Aravena2016_cii}, wherein the fidelity of blindly detected [CII] line candidates was expressed as a function of SNR only, given the limited statistical strength, i.e., noise samplings per SNR bin, to test the dependence of fidelity against line width in the Pilot program. As argued in \citet{GL2019_3mm}, SNR alone is insufficient to provide an accurate estimate of fidelity in cases where the data potentially contains emission lines of varying widths, as in ASPECS. Line candidates detected with the same SNR for different spectral convolutions of the data will have different overall significance (not captured by SNR) that depends on the number of independent elements (i.e., frequency channels) in a given convolution. This effect is manifest in the ASPECS 1.2~mm blind line search, and has been presented in \citet{Decarli2020_1mm}, where the authors found that broader line candidates tend to have higher fidelity than narrower line candidates, at a given SNR (see top panel of their Figure 2). 

Only five [CII] line candidates were returned by the blind search performed in \citet{Decarli2020_1mm}, which yielded a catalog containing a total of 234 line candidates with fidelity~$>0.2$. All five of the [CII] candidates are modest SNR ($=5$--6) and low fidelity ($< 0.8$) detections, with four out the five candidates characterized by fidelity $< 0.5$. We note that the SNR~$=5$--6 range reflects the threshold where fidelity rapidly decreases to zero (Figure 2, \citet{Decarli2020_1mm}). Only one [CII] line candidate is considered a good match\footnote{Here, we require that the [CII] line candidate at $z_{\ciimath}$ and the known source at $z$ have (1) a spatial offset within 0.1~arcsec and (2) a redshift separation $(z - z_{\ciimath})/(1+z) < 0.1$.} to a known optical/near-IR counterpart with a photometric redshift, but, based on the analysis above, the probability that the line is spurious is $>70$\% (i.e., fidelity~$= 0.28$). We therefore discount it, along with the remaining [CII] line candidates, upon inspecting their Band 6 spectra and continuum postage stamps.

Results 

\subsection{Continuum emission: Individual sources and stack} \label{sec:lae_cont}

\begin{figure}
\gridline{\fig{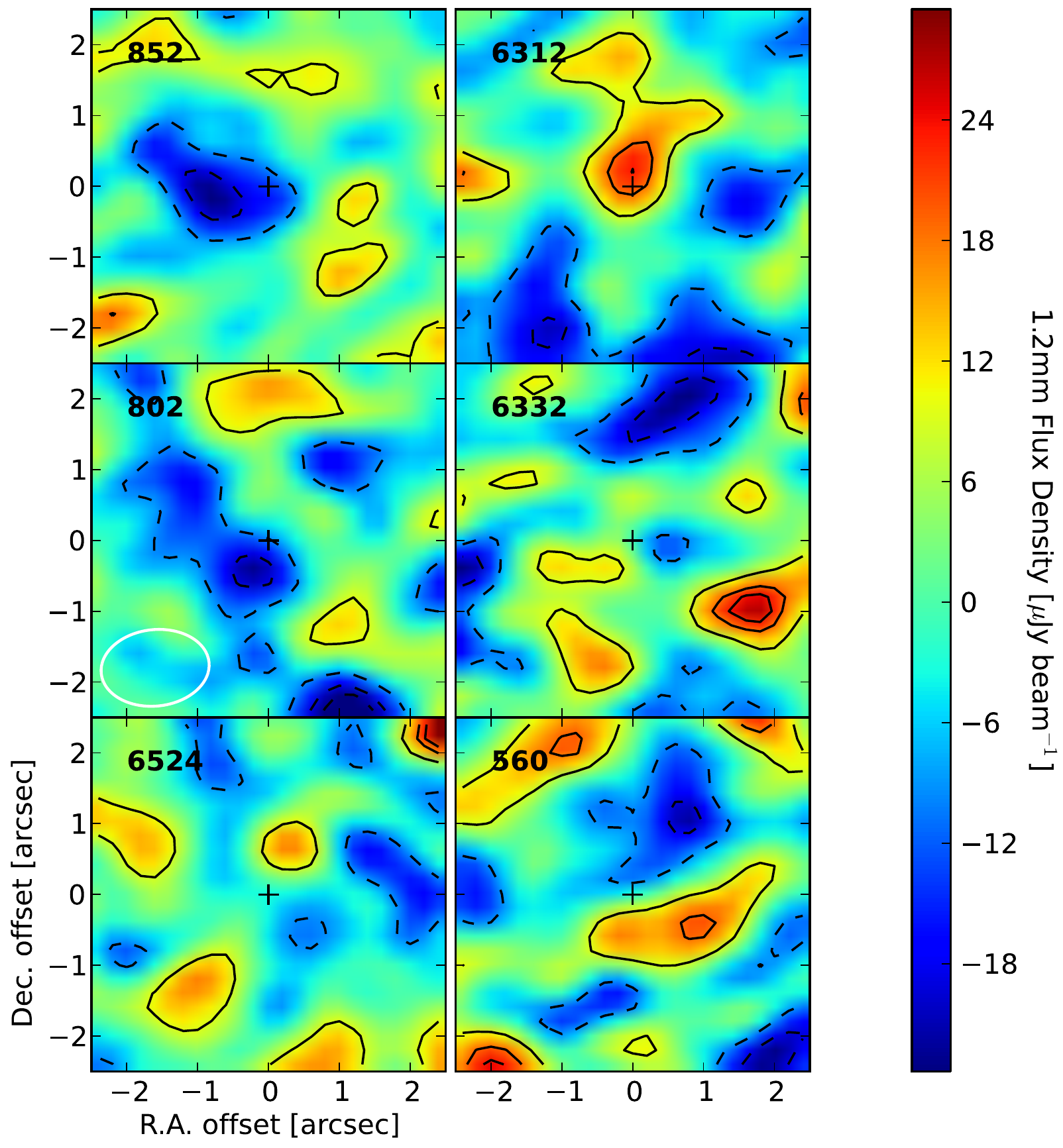}{0.45\textwidth}{(a)}}
\gridline{\fig{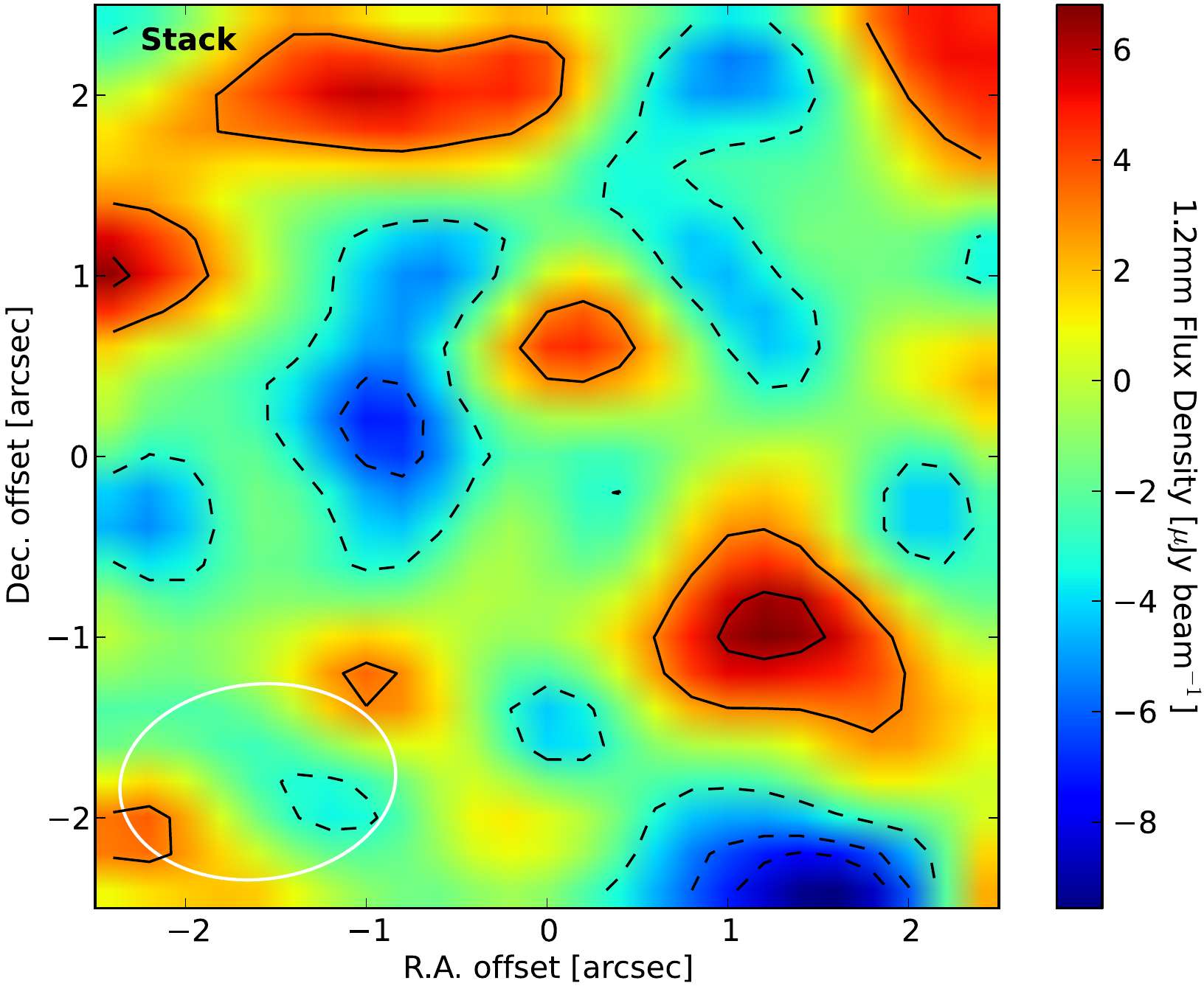}{0.45\textwidth}{(b)}}
\caption{\textbf{(a)}: $5'' \times 5''$ postage stamps of 1.2~mm continuum images (without PB corretion) centered at locations of MUSE LAEs in ASPECS. The LAE position in each cutout is indicated with ``+" symbol; refer to Table~\ref{tab:lae_props} for sky coordinates. \textbf{(b)}: Mean 1.2~mm continuum emission obtained by stacking on the locations of the six MUSE LAEs shown in the lefthand panel. Contours in both panels represent the emission at $1.0\sigma$ ($=9.3\mu$Jy~beam$^{-1}$), $2.0\sigma$, and $2.5\sigma$ levels, with dashed and solid contours corresponding to negative and positive flux densities, respectively. For reference, the synthesized beam is shown as an ellipse in the bottom left corners of the postage stamp for MUSE 802 and the stack.}
          \label{fig:lae_cont}
\end{figure}

In addition to extracting spectra for the MUSE LAEs described in Section~\ref{sec:lae}, we have also searched for the presence of continuum emission at the corresponding locations in the line-free 1.2~mm map obtained in \citet{GL2020_1mm}. Continuum image cutouts ($5'' \times 5''$; no primary beam correction) centered at individual LAE positions are shown in Figure~\ref{fig:lae_cont}(a), with signal-to-noise contours overlaid after adopting an rms value of 9.3~$\mu$Jy per beam \citep{GL2020_1mm}. MUSE~6312 is the only source with plausible continuum emission, observed at the 2--2.5$\sigma$ level, but improved sensitivity is needed to assert the reality of this emission. We can improve our sensitivity on the average 1.2~mm continuum emission for all the LAEs in our sample by stacking. In Figure~\ref{fig:lae_cont}(b), we show the results of a continuum stack on the central location of the six MUSE LAEs, generated by averaging the emission in continuum images and weighting each pixel by the mosaic sensitivity pattern. The stacked continuum image has an rms noise equal to 2.98~$\mu$Jy~beam$^{-1}$. Non-detections of 1.2~mm continuum flux in individual LAEs with comparable L$_{\mathrm{Ly}\alpha}$ ($\lesssim 5\times 10^{42}$~erg~s$^{-1}$) at similar redshift have been previously reported down to $\sim10$-15~$\mu$Jy ($1\sigma$) \citep[e.g.,][]{Knudsen2016, Bradac2017}.

We can use the non-detection of continuum in the stack to place an upper limit on the dust-obscured SFR in the LAEs. Adopting a dust temperature of 30~K (50~K) and emissivity index $\beta=1.6$, we integrate a modified black-body spectrum across the far-infrared (FIR) wavelengths 42.5--122.5~$\mu$m \citep{Helou1988} to place an upper limit ($3\sigma$) on the FIR luminosity $L_{\mathrm{FIR}} < 2.7\times10^9$~L$_{\odot}$ ($1.2\times10^{10}$~L$_{\odot}$) for these sources.\footnote{While the warmer CMB temperatures at the redshifts relevant to this analysis can reduce detectability of the intrinsic continuum flux density or provide additional dust heating \citep{daCunha2013}, we follow the reasoning in, e.g., \citet{Willott2015}, and argue that---with a single flux density measurement---these competing effects of the CMB background are poorly constrained.} (We have adopted an emissivity index and dust temperatures consistent with findings for $z\sim5.5$ galaxies in \citet{Faisst2020}.) Using the conversion $\mathrm{SFR} / L_{\mathrm{FIR}} = 1.5\times10^{-10}$~M$_{\odot}$~yr$^{-1}$~L$_{\odot}^{-1}$ (for Chabrier IMF, as in \citet{CarilliWalter2013}), we find IR-based $\mathrm{SFR}< 0.4$~M$_{\odot}$~yr$^{-1}$ (1.8~M$_{\odot}$~yr$^{-1}$) . 

Low to negligible levels of obscured star formation in the MUSE LAEs are consistent with results from a related independent study of 1.2~mm continuum emission from $\sim1400$ galaxies with either Lyman break or photometric redshift-selection at $z=1.5$--10 in the ASPECS LP 4.2~arcmin$^2$ footprint \citep{Bouwens2020}. The authors there---and in the ASPECS 1.2~mm continuum source blind search \citep{GL2020_1mm}---do not report any continuum detections beyond $z=4$ for their sample of UV-selected galaxies, which includes the \citet{Bouwens2015} LBGs used in this work; we can confirm non-detections for these overlapping sources after examining the corresponding 1.2~mm continuum postage stamps. \citep{Bouwens2020} also searched for 1.2~mm continuum flux in a stack of low mass ($<10^{9.25}$~M$_{\odot}$) galaxies across the full redshift range probed by their sample, finding an average 1.2~mm continuum flux density of $-0.1\pm0.4$~$\mu$m for the 1,253 galaxies in the stack, implying that the obscured SFR in these galaxies is approximately zero (assuming $z=4$ for the entire stack). 
 
\section{Discussion}
\label{sec:discussion}

In this section, we compare our findings in Sections~\ref{sec:optical_nir}--\ref{sec:blind} to the results of the [CII] line search in the ASPECS Pilot study \citep{Aravena2016_cii}, and place the findings of the targeted [CII] searches on LAEs and LBGs in ASPECS (Sections~\ref{sec:lae} and \ref{sec:lbg}) in the context of empirical and predicted $L_{\ciimath}$-SFR relations at $z\sim0$ and $6 \leq z \leq 8$. Finally, we use the absence of reliable detections in the blind search to place upper limits on the cumulative [CII] source densities. 

\subsection{Comparison to ASPECS Pilot}

\citet{Aravena2016_cii} presented the results of a [CII] line search in the ASPECS Pilot survey \citep{Walter2016_survey}. As the precursor to the Large Program, ASPECS Pilot shared the same survey strategy (e.g., array configuration, frequency setups, comparable survey depth) as ASPECS LP,  but targeted a smaller, 1~arcmin$^2$ patch of sky within the UDF. The final mean rms noise achieved was 0.42~mJy beam$^{-1}$ per 62.5~MHz channel, roughly constant across the survey bandwidth, which is a factor 1.4 higher than the mean rms of 0.30~mJy~beam$^{-1}$ per 62.5~MHz channel for ASPECS LP due to more favorable weather conditions in the latter campaign. 

The blind line search in the lower sensitivity ASPECS Pilot data cube returned 14 [CII] line candidates down to lower signal-to-noise threshold ($\sim4.5)$ than considered in the LP blind search. Fidelities were obtained in a similar probabilistic fashion as expressed in Equation~\ref{eq:fidelity_LP}, but lacking the dependence on line width: $\mathrm{fidelity} = 1 - N_{neg}\left(\mathrm{SNR})\right)/N_{pos}\left(\mathrm{SNR})\right)$. Two of these candidates were not associated with known nearby optical dropout galaxies (at any redshift), and were presented based on an assessment of their fidelities ($>70$\%), with the caveat that one line candidate overlapped with an atmospheric absorption feature, and that further ALMA spectroscopy would be needed to confirm the candidates' reality. The remaining 12 candidates were found after applying positional priors---set by their proximity ($<1.0$~arcsec) to optical dropout galaxies at $5.5 < z < 8.5$---to the blindly detected line candidates, and setting a lower threshold to the fidelity ($>40$\%), given the associations with optical counterparts. 
 
Because 13 of the 14 [CII] line candidates lie within the ASPECS LP HPBW, we extract their spectra from the peak pixel position reported in \citet{Aravena2016_cii} (see their Table~2) in  the LP data cube to independently confirm or reject the presence of line emission.\footnote{One source in \citet{Aravena2016_cii}, ID30, lies in a region where the LP mosaic sensitivity is $\sim30$\%, so we do not formally include it in the results of this study, after visually inspecting the data.} The new observations disprove the reality of all 13 candidates. These non-detections reinforce the fact that even line candidates with optical associations at the $4.5 < \mathrm{SNR} < 5.5$ level can be mistaken for real sources, and motivate (1) the development of improved techniques for assessing the line fidelity, and (2) the adoption of a more conservative approach (i.e., setting higher flux thresholds) when presenting line candidates, as described, e.g., in Section~\ref{sec:blind}, and references therein.

\subsection{L$_{\ciimath}$-SFR relation at $6 \leq z \leq 8$}

Figure~\ref{fig:lcii_sfr} shows the upper limits ($5\sigma$) on $L_{\ciimath}$ measured for MUSE LAEs and LBGs in ASPECS as a function of their inferred total SFRs, including unobscured and, where available, obscured SFR contributions.  For comparison, we include literature data for [CII] observations in the redshift range $z=6$--7 compiled by \citet{Matthee2019}, wherein the authors re-calculated SFRs for the entire sample in a consistent manner, setting a standard IMF (Salpeter) and dust temperature (45~K) for all galaxies. For consistency with our derived SFRs, which assumed a Chabrier IMF, we multiply their UV- and IR-based SFR values by IMF conversion factors equal to 0.63 and 0.87, respectively \citep{MD2014}. 

Here, the depicted error on SFR for the LAE sample includes the uncertainty associated with the SED fitting, as well as a 5$\sigma$ upper limit on dust-obscured SFR ($= 7.6$--10.2~M$_{\odot}$~yr$^{-1}$) derived from the extrapolation of total IR luminosity ($L_{IR}$[8--1000~$\mu$m]) from the 1.2~mm continuum data. To be consistent with the plotted literature data points, we adopt the same MBB parameters and $L_{IR}$-SFR conversion factor as in \citet{Matthee2019} when estimating $L_{IR}$ and the SFR. (Note that the MBB parameters used in \citet{Matthee2019} differ slightly from the chosen parameters in Section~\ref{sec:lae_cont}.\footnote{Specifically, instead of our adopted dust temperatures at 30~K and 50~K and $\beta=1.6$, \citet{Matthee2019} adopt a single 45~K dust temperature and $\beta$=1.5. As an example comparison, the latter parameterization results in roughly $0.65$ times lower IR luminosity than the MBB with 50~k dust temperature and $\beta=1.6$, when integrating over the same wavelength range from 8~$\mu$m--1000~$\mu$m.}) We do not show individual uncertainties on SFR for the LBG sample; the uncertainties are comparable to what was found for the MUSE LAEs. 

For sources with SFRs less than a few M$_{\odot}$~yr$^{-1}$, the ASPECS LP non-detections are unsurprising based on the $L_{\ciimath}$-SFR relation calibrated for local galaxies \citep{DeLooze2014}. In the literature, the only [CII] detections reported in this SFR regime are targeted ALMA observations of an LBG \citep{Knudsen2016} and LAE \citep{Bradac2017} where emission has been magnified by strong gravitational lensing.\footnote{After correcting for a lensing magnification factor of 5, the LAE in \citet{Bradac2017} has $L_{\mathrm{Ly}\alpha} = 1.3\times10^{42}$~erg~s$^{-1}$, which is comparable to the range of Lyman-$\alpha$ luminosities probed by the MUSE LAEs in ASPECS.} 

There are, however, a few LBGs within the survey HPBW where the ASPECS [CII] detection threshold is more constraining. Explicitly, if we set for ASPECS the mean 5$\sigma$ survey depth on SFR based on the locally calibrated \citet{DeLooze2014} $L_{\ciimath}$-SFR relation (using their``star-forming HII region/starburst'' sample), then detections might have been expected for galaxies with SFR $ \geq 16\pm8$~M$_{\odot}$~yr$^{-1}$ (SFR $\geq 32\pm18$~M$_{\odot}$~yr$^{-1}$) at $z = 6$ ($z = 8$), where the error bars reflect the uncertainty in the $L_{\ciimath}$-SFR relation. We motivate the choice of the local $L_{\ciimath}$-SFR relation to set a fiducial survey depth in SFR in light of results of Schaerer et al. 2020, who find little to no evolution in the local $L_{\ciimath}$-SFR relation since $z \leq 6$,  because the consensus at $z>6$ on the nature (e.g., its slope, scatter, and linearity) of this relation has not converged. 

For instance, the $5\sigma$ upper limit on [CII] luminosity in XDFI-2374646327---the LBG in ASPECS with the highest observed SFR ($\sim50$~M$_{\odot}$~yr$^{-1}$)---is $L_{\ciimath} < 2.0\times10^8$~L$_{\odot}$, which is more than three times lower than expected from the best-fit relation for local star-forming galaxies per \citet{DeLooze2014}, and is also below the observed 0.27~dex scatter in the \citet{DeLooze2014} relation (gray-shaded band in Figure~\ref{fig:lcii_sfr}). The limit is consistent, however, with the locally-calibrated $L_{\ciimath}$-SFR relation in \citet{Diaz2017, diaz2013}, where a turnover in $L_{\ciimath}$ is observed for galaxies with $\mathrm{SFR} \gtrsim 30$~M$_{\odot}$~yr$^{-1}$. The discrepancy between the two locally calibrated relations is partly explained by the fact that the SFR surface densities ($\sim85$~M$_{\odot}$~yr$^{-1}$~kpc$^{-2}$) probed in the GOALS sample of \citet{Diaz2017, diaz2013} are nearly 2--3 times higher than in the objects compiled by \citet{DeLooze2014} and other local galaxy samples (e.g., \citet{HerreraCamus2015}). Thus, one possible explanation for the [CII] deficiency in this source could be the presence of high surface density of star formation, which is supported, e.g., by findings in \citet{Ferrara2019_ciisfr}, who predict a deficiency in [CII] luminosity surface density with respect to the \citet{DeLooze2014} relation for galaxies at $z>5$ with SFR surface densities above $\sim85$~M$_{\odot}$~yr$^{-1}$~kpc$^{-2}$. Other factors, such as metallicity, as proposed, e.g., in \citet{Vallini2015}, might also play a role, though we note that a saturation of [CII] emission in high SFR surface density systems is predicted to be dominant over the effects of metallicity \citep{Ferrara2019_ciisfr}.

Excluding XDFI-2374646327, then the derived upper limits for the remaining $\mathrm{SFR} \gtrsim 10$~M$_{\odot}$~yr$^{-1}$ sources in ASPECS are consistent with the observed scatter in previous targeted ALMA observations, as well as the local $L_{\ciimath}$-SFR relations calibrated by \citet{diaz2013} and \citet{DeLooze2014}. 

\subsection{Cumulative [CII] source densities at  $6 \leq z \leq 8$}

\begin{figure}
    \centering
\includegraphics[width=0.45\textwidth]{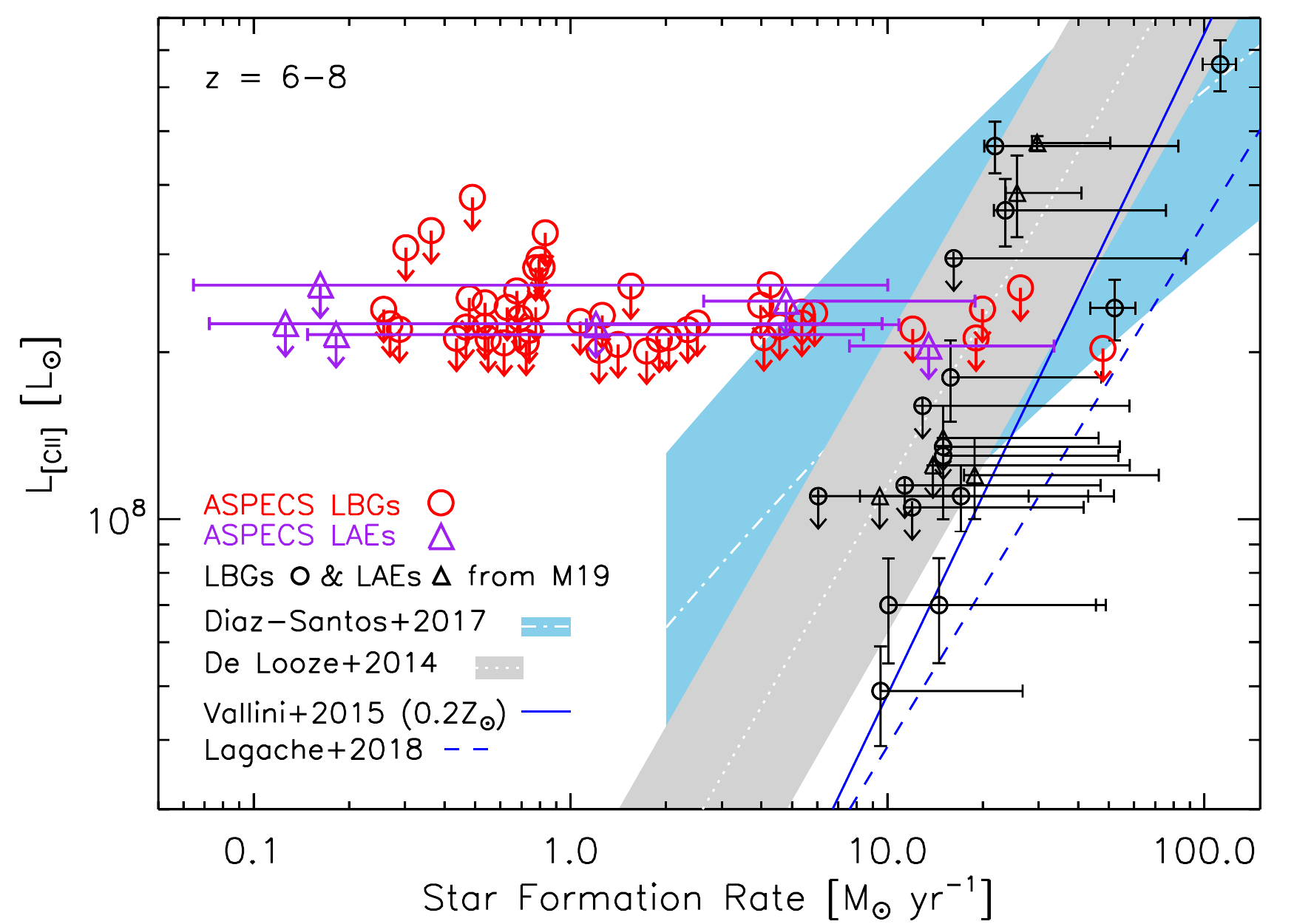}
        \caption{$L_{\ciimath}$-SFR relation at $z>6$, with data points distinguished by target selection: circle and triangle symbols reflect galaxies initially discovered as LBGs and LAEs, respectively. 5$\sigma$ upper limits on $L_{\ciimath}$ are presented for LBGs (red circles) and MUSE LAEs (purple triangles) within the ASPECS HPBW and redshift coverage. Detections and 5$\sigma$ upper limits from the literature, as compiled by \citet{Matthee2019}, are shown for comparison as black symbols. Also shown are the local $L_{\ciimath}$-SFR relations calibrated by \citet{DeLooze2014} for HII region/star-forming systems, including starbursts (dark gray band; 1$\sigma$ dispersion = 0.27 dex), and by \citet{diaz2013} for (U)LIRGs (light blue band; 1$\sigma$); white dotted lines within each band indicate the best-fit relations for each sample. Theoretical predictions for the $L_{\ciimath}$-SFR relation at $z=6$--$7$ are plotted from \citet{Vallini2015} and \citet{Lagache2018} (solid and dashed blue linestyles, respectively).}
          \label{fig:lcii_sfr}
\end{figure}

\begin{figure*}
    \centering
    \includegraphics[width=1.0\textwidth]{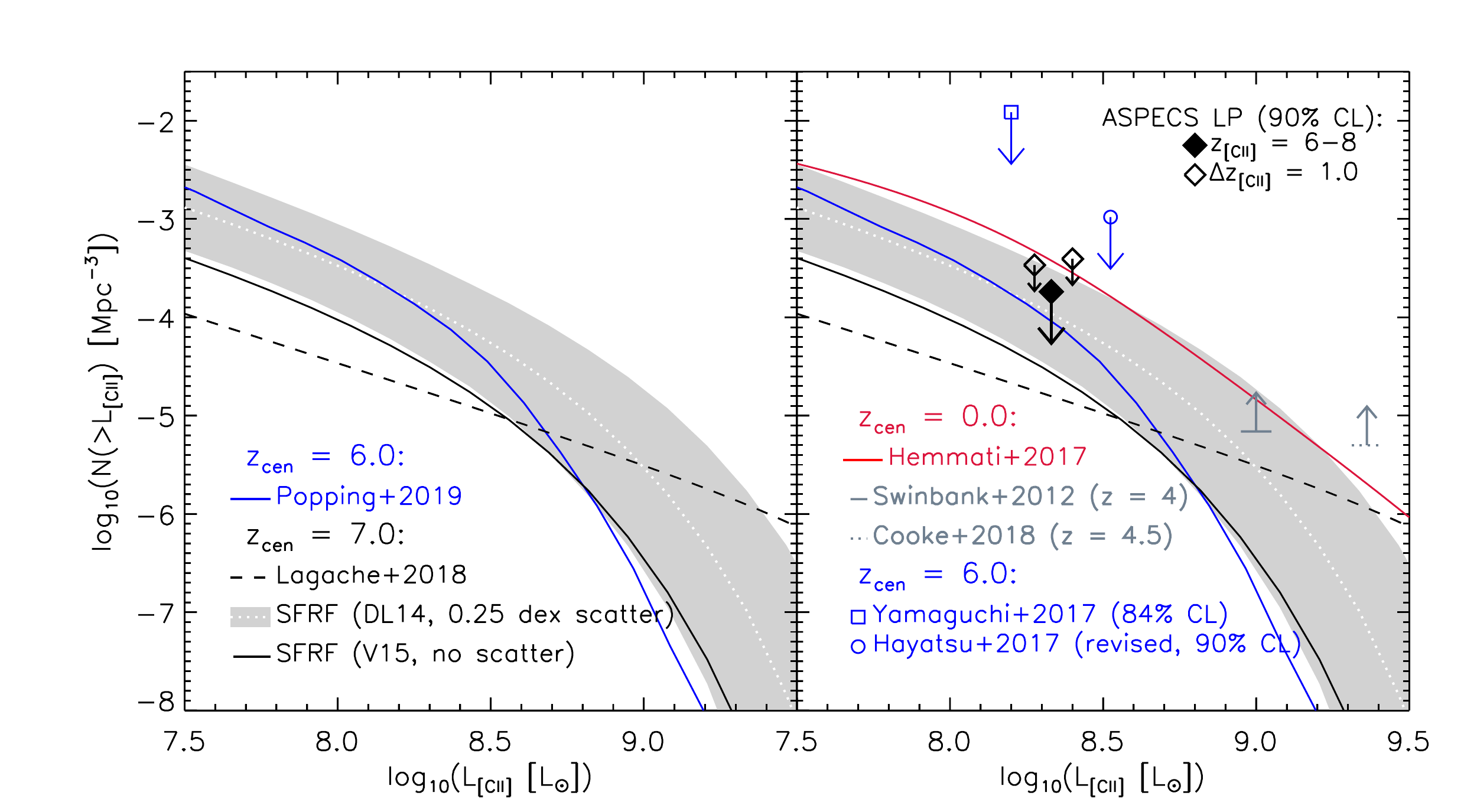}
    \caption{(\emph{left}:) Theoretical predictions for $N(>L_{\ciimath})$ vs. $L_{\ciimath}$ at $z=6$ (blue curve: \citet{Popping2019}) and $z=7$ (white dotted curve: best fit for simulated number densities based on the \citet{Smit2016} SFRF convolved with $L_{\ciimath}$-SFR relation from the \citet{DeLooze2014} HII region/starburst calibration, including a 0.25~dex scatter on $L_{\ciimath}$-SFR shown as the gray band; black solid curve: same as white dotted curve, except using $L_{\ciimath}$-SFR relation from \citet{Vallini2015} (0.2~$Z_{\odot}$); black dashed curve: \citet{Lagache2018}). (\emph{right:}) Observational constraints on $N(>L_{\ciimath})$ vs. $L_{\ciimath}$, with upper limits (90\% confidence level) derived from the full ASPECS survey volume covering $z=6$--8 (filled black diamond) and the literature at $z\sim6$ \citep{Yamaguchi2017, Hayatsu2017}. The [CII] cumulative number densities measured at $z\sim0$ \citep{Hemmati2017}, and lower limits at $z\sim4$--5 \citep{Swinbank2012, Cooke2018} are plotted as the gray arrows. Theoretical predictions from the left panel are underlaid to facilitate comparison.}
    \label{fig:nintcum_cii}
\end{figure*}

Figure~\ref{fig:nintcum_cii} shows the upper limits (downward-pointing arrows) on the cumulative [CII] source densities (i.e., number density of [CII] emitters with luminosity greater than $L_{\ciimath}$) derived from the ASPECS LP blind search. The choice of presenting cumulative [CII] source densities is for consistency with \citet{Aravena2016_cii}; Figure~\ref{fig:nintcum_cii}, and also Figure~\ref{fig:lcii_sfr} from Section~4.2, replace the results of that work.  We present limits at 90\% confidence level assuming Poisson statistics for zero detections \citep{Gehrels1986} for the full survey volume spanning $z=6$--8, as well as for smaller volumes corresponding to redshift ranges $z=6$--7 and $z=7$--8.\footnote{Per \citet{Gehrels1986}, the 1.0$\sigma$, 1.3$\sigma$, 2.0$\sigma$, and 3.0$\sigma$ limits in Gaussian statistics correspond to single-sided Poissonian upper limits at confidence levels of 84.1\%, 90.0\%, 97.5\%, and 99.9\%, respectively. The Gaussian 3$\sigma$ upper limit, e.g., can be derived for ASPECS by multiplying the upper limit for zero detections quoted at 90\% confidence level by a factor 2.869.}  The ASPECS [CIl] luminosity depths (rightward-pointing arrows) are the 5$\sigma$ upper limits derived from the average RMS per channel in the data cube across the relevant frequency ranges for each bin\footnote{after adopting the central redshift corresponding to each redshift bin} (Figure~\ref{fig:ciilumlim}). Table~\ref{tab:nintcum} summarizes the measured limits for the different redshift ranges. We note that \citet{Decarli2020_1mm} present upper limits on the [CII] luminosity function (in units of Mpc$^{-3}$~dex$^{-1}$) in ASPECS 1.2~mm data, derived using the blind search algorithm developed there (and in references therein) for CO and other lines within the ASPECS survey bandwidth. To facilitate comparison, we convert the $3\sigma$ limits on the luminosity function presented in their Table~4 to the appropriate number of Gaussian $\sigma$ equivalent to the 90\% confidence level ($\approx1.3\sigma$), and integrate the resulting limits to arrive at a cumulative number density of $< 1.94\times10^{-4}$~Mpc$^{-3}$ for [CII] emitters with luminosities greater than their lowest luminosity bin centered at $L^{\prime}_{\ciimath} = 1.26\times10^9$~K~km~s$^{-1}$~pc$^2$, or $L_{\ciimath} = 2.77\times10^8$~L$_{\odot}$.

\begin{table}[b]
    \centering
    \caption{Limits on cumulative [CII] source densities}
    \begin{tabular}{c c c c}
    \hline \hline 
    $z$ range & $n(>L_{\ciimath})$ & $>L_{\ciimath}$ & Ref. \\
                            &   [Mpc$^{-3}$]        & [L$_{\odot}$]    &   \\
           (1)             &     (2)                      &   (3)          & (4)      \\
    \hline
    $6$--$8$  &   $<1.82 \ (5.22) \times10^{-4}$  & $>2.14\times10^8$ & U21 \\
    $6$--$8$  & $<1.94 \ (4.47) \times10^{-4}$  & $>2.77\times10^8$ &  D20 \\
    $6$--$7$   & $<3.40 \ (9.75) \times10^{-4}$   & $>1.89\times10^8$ & U21 \\
    $7$--$8$   & $<3.93 \ (11.3) \times10^{-4}$   & $>2.51\times10^8$ & U21 \\
    \hline
    \multicolumn{4}{p{1.0\columnwidth}}{---\emph{Notes:} (1) Redshift range (2) Upper limit at 90\% (99.9\%) confidence level on cumulative number densities. For D20, tabulated $3\sigma$ upper limits on the [CII] luminosity function (in units of Mpc$^{-3}$~dex$^{-1}$) have been converted to $1.3\sigma$ upper limits in each luminosity bin to derive an equivalent one-sided Poissonian limit on the integrated, cumulative [CII] source densities at 90\% confidence level. No conversions on the upper limits in D20 were performed in the case of the Poissonian upper limit quoted at 99.9\% confidence level, which corresponds to the $3\sigma$ limit in Gaussian statistics. (3) [CII] luminosity depth ($5\sigma$). (4) Reference (U21: This work; D20: \citet{Decarli2020_1mm})}
    \end{tabular}
    \label{tab:nintcum}
\end{table}

We show for comparison the previous observational constraints at $z\sim6$ in the literature from \citet{Hayatsu2017}\footnote{The authors in \citet{Hayatsu2019} show the detections in \citet{Hayatsu2017} to be spurious, so we have re-measured upper limits (90\% confidence level) for \citet{Hayatsu2017} adopting a mean RMS noise 0.8~mJy~beam$^{-1}$ per 36~km~s$^{-1}$ channel and survey volume $2.2\times10^3$~Mpc$^3$.} and \citet{Yamaguchi2017}. The [CII] number counts observed for local galaxies \citep{Hemmati2017} are also shown (solid red curve). The ASPECS limits on the cumulative source densities imply that the [CII] number density is at least a factor of 2 lower than measured at $z\sim0$ at the ASPECS 5$\sigma$ $L_{\ciimath}$ depth of $2.14\times10^8$~$L_{\odot}$. At intermediate redshifts $z\sim4$--6, the ALPINE survey also provides constraints on [CII] number counts \citep{Yan2020, Loiacono2020}, though a direct comparison with ASPECS results, or any [CII] number counts that originate from flux-limited surveys (like the GOALS+RBGS sample in \citet{Hemmati2017}), is complicated by the nature of the target selection \citep{LeFevre2019} in that survey sample. 

The cumulative [CII] source densities can indicate whether the current understanding of the $L_{\ciimath}$-SFR relation at high redshift established by previous targeted ALMA studies of individual optically-selected sources is consistent with the results of the ASPECS [CII] blind search. The white dotted curve in Figure~\ref{fig:nintcum_cii}) represents the best fit to simulated [CII] number counts at $z=7$, generated using a Monte Carlo simulation that predicts the number density of [CII] emitters at a given $L_{\ciimath}$ and SFR by sampling the star-formation rate function (SFRF, in units of M$_{\odot}$~yr$^{-1}$~Mpc$^{-3}$) measured in GOODS-N and -S from \citet{Smit2016} and applying the $L_{\ciimath}$-SFR relation for galaxies from \citet{DeLooze2014}. The simulations account for uncorrelated errors in the SFRF Schechter parameters and include a 0.25~dex scatter in $L_{\ciimath}$-SFR (gray band in Figure~\ref{fig:lcii_sfr}); the inclusion of correlated errors in the SFRF Schechter parameters would decrease the overall dispersion in the simulations (gray swath in Figure~\ref{fig:nintcum_cii}; 1$\sigma$). Note that the UV-based SFRFs have been corrected for dust extinction using an SMC-like attenuation law; the actual attenuation law is more likely between the SMC curve and a Calzetti law \citep{Bouwens2020}. Our deepest upper limit (90\% confidence level) on the cumulative number density (centered at $z=7$) does not rule out the best fit to this simulated model, but begins to place useful constraints on the predicted scatter, driven largely by the dispersion in $L_{\ciimath}$-SFR; it is important to note, however, that the reported upper limit in this work neglects the effect of cosmic variance. If we replace the \citet{DeLooze2014} $L_{\ciimath}$-SFR relation in our simulation with the prescription found in \citet{Vallini2015}, then the resulting average fit is the solid black curve in the same Figure, which is roughly 1~dex below our upper limit. Thus, while the blind search is unbiased and could potentially reveal a population of [CII] emitters that are not identified via optical selections, our results in Figure~\ref{fig:nintcum_cii} indicate that the [CII] number density at $6\leq z \leq 8$ in UDF is broadly consistent with expectations based on the current understanding obtained by targeted observations of LAEs and LBGs (Figure~\ref{fig:lcii_sfr}).

Additional theoretical models at $z=6$ \citep{Popping2019} and $z=7$ \citep{Lagache2018} are also plotted, for comparison (Figure~\ref{fig:nintcum_cii} (left panel)).

\subsection{Cosmic molecular gas mass density}

Using [CII] as a molecular gas tracer \citep[][]{Zanella2018}, we complement the recent measurements of cosmic molecular gas density, $\rho_{\mathrm{H}2}$, presented in the ASPECS study by \citet{Decarli2020_1mm} (their Figure~9), extending these constraints to $z > 6$. At this redshift, the COLDz survey, which targeted CO emission from 30--39~GHz with the VLA, places the only existing constraints on $\rho_{\mathrm{H}2}$ \citep{Riechers2019}. 

We here use, for the purpose of a rough estimate, the $L_{\ciimath}$-H$_2$ conversion factor, $\alpha_{\ciimath} = 31$~M$_{\odot}$~L$_{\odot}^{-1}$, empirically calibrated in \citet{Zanella2018} to guide our estimate of $\rho_{\mathrm{H}2}$. We refer the interested reader to that work (and references therein) for thorough discussion of related caveats on the reliability of [CII] as a molecular gas tracer, e.g., its prevalence in different ISM phases. This value for $\alpha_{\ciimath}$ appears to be invariant (within a scatter of 0.3~dex) across the different samples explored in their work, including local main-sequence (MS) and starbursting galaxies, low-metallicity local dwarfs, and high redshift ($z\sim2$--5.5) MS and starburst galaxies.  

Per our deepest constraints on the total number density of [CII] emitters for the $z = 6$--8 redshift bin (Table~\ref{tab:nintcum}), the [CII] luminosity density for all galaxies above our 5$\sigma$ depth in $L_{\ciimath}$ at this redshift cannot exceed $(2.14\times10^8$~L$_{\odot}$) $\times \ (1.82\times10^{-4}$~Mpc$^{-3}$) $=3.89\times10^4$~L$_{\odot}$~Mpc$^{-3}$. This implies that $\rho_{\mathrm{H}2}$ from galaxies with $L_{\ciimath}>2.14\times10^8$~$L_{\odot}$ cannot exceed $\alpha_{\ciimath}\times(3.89\times10^4$~L$_{\odot}$~Mpc$^{-3}$) $=1.2\times10^6$~M$_{\odot}$~Mpc$^{-3}$. Comparing with constraints on $\rho_{\mathrm{H}2}$ from the COLDz survey \citep{Riechers2019}, we find that our upper limit sits just below the measured range of their CO-derived estimate of $\rho_{\mathrm{H}2}=0.14$--$1.1\times10^7$~M$_{\odot}$~Mpc$^{-3}$ in the $z = 4.9$--6.7 redshift bin. While assumptions regarding, e.g., the CO-H$_2$ conversion factor outlined in \citet{Riechers2019} and the uncertain nature of the $\alpha_{\ciimath}$ factor applied in the context of this work are likely dominant sources of this discrepancy, we point out that (1) the COLDz measurement reflects contributions from lower redshift galaxies than in ASPECS, and (2) there might be non-negligible contributions to the [CII] luminosity density, and thus $\rho_{\mathrm{H}2}$, from lower luminosity [CII] emitters, depending on the faint-end slope of the [CII] luminosity function at $z\sim7$.


\section{Conclusions}
\label{sec:conclusions}

We present a targeted search for [CII] emission from optically-selected galaxies within the ASPECS LP 1.2~mm data cube ($\nu_{obs} = 212$--272~GHz), as well as the deepest constraints on the number density of [CII] emitters at the end of Reionization, from $z=6$--8. Key results include the following:

\begin{enumerate}[label=(\roman*)]
\item With a mean RMS sensitivity 0.30~mJy~beam$^{-1}$ per 62.5~MHz channel across the full ASPECS B6 bandwith---corresponding to average 5$\sigma$ depths in $L_{\ciimath}=2.14\times10^8$~L$_{\odot}$ and $\mathrm{SFR} \sim 20~M_{\odot}$~yr$^{-1}$ (per \citet{DeLooze2014})---we place upper limits on [CII] line luminosity for 6 LAEs and 45 LBGs within the ASPECS HPBW (2.9~arcmin$^2$). For these sources, the derived upper limits are consistent with previous targeted ALMA observations of $z=6$--7 LAEs and LBGs, as well as the local $L_{\ciimath}$-SFR relations from \citet{DeLooze2014} or, in the case of a single LBG with estimated SED-based SFR $\sim 50$~M$_{\odot}$~yr$^{-1}$, from \citet{Diaz2017}.
\item Upon stacking the 1.2~mm continuum data for the 6 LAEs in our survey field, we can probe emission down to an RMS noise level equal to 2.98~$\mu$Jy~beam$^{-1}$. Adopting a template modified black body spectrum with a dust temperature of 50~K, $\beta$=1.6, and integrating from FIR wavelengths 42.5--122.5~$\mu$m, we place a $3\sigma$ upper limit on $L_{\mathrm{FIR}} < 1.2\times10^{10}$ and $\mathrm{SFR} < 1.8$~M$_{\odot}$~yr$^{-1}$ (or  $L_{\mathrm{FIR}} < 2.7\times10^{9}$ and $\mathrm{SFR} < 0.4$~M$_{\odot}$~yr$^{-1}$ for dust temperature of 30~K).
\item In a volume of $\sim12,500$~comoving~Mpc$^{3}$, we find that the number density of [CII] emitters with line luminosity greater than $2.14\times10^8$~L$_{\odot}$ in the redshift range $z=6$--8 is less than $1.82\times10^{-3}$~Mpc$^{-3}$ (90\% confidence level), consistent with results in \citet{Decarli2020_1mm}, who performed a blind line search for all spectral lines within ASPECS spectral coverage. Our upper limits indicate evolution of the [CII] LF from $z=6$--8 to $z=0$ at the quoted [CII] depth for ASPECS. 
\end{enumerate}

Looking forward, there are different avenues to make further progress in this field. One promising avenue would be to obtain significantly deeper [CII] and dust continuum observations of individual reionization sources that are selected using various techniques (including the ones used in this paper, i.e., Lyman break galaxies and Lyman alpha emitters). This approach is well within the reach of dedicated observations with ALMA---including recent ALMA Large Programs REBELS (PI: R. Bouwens) and the ALMA Lensing Cluster survey (PI: K. Kohno)---and will be complemented in the future with new systemic redshift measurements from JWST. Statistical approaches applied to existing ALMA datasets, such as the power spectrum analysis (as demonstrated in \citet{Uzgil2019} and \citet{Keating2020}) provide efficient tools for probing low fidelity emission from faint galaxies below survey detection thresholds. Line intensity mapping datasets---tailored to this statistical approach---will also be available from ongoing and future experiments using large field of view instruments on single dish telescopes (e.g., EoR-Spec \citep{Cothard2020}; SuperSpec \citep{Karkare2020,Redford2018}; CONCERTO \citep{Lagache2018}; TIME \citep{Sun2020}) to probe aggregate [CII] emission at the highest redshifts. 
\newline

The authors thank the anonymous referee for a constructive report. We thank Takuya Hashimoto for sharing equivalent width data for the MUSE LAEs in ASPECS. BDU would like to thank Andrea Ferrara and the cosmology group at Scuola Normale Superiore, where part of this work was done, for their hospitality and useful discussions. DR acknowledges support from the National Science Foundation under grant numbers AST-1614213 and AST-1910107. DR also acknowledges support from the Alexander von Humboldt Foundation through a Humboldt Research Fellowship for Experienced Researchers. TD-S acknowledges support from the CASSACA and CONICYT fund CAS-CONICYT Call 2018. HI acknowledges support from JSPS KAKENHI Grant Number JP19K23462.
 
 \emph{Facility:} ALMA data: 2016.1.00324.L. ALMA is a partnership of ESO (representing its member states), NSF (USA) and NINS (Japan), together with NRC (Canada), NSC and ASIAA (Taiwan), and KASI (Re- public of Korea), in cooperation with the Republic of Chile. The Joint ALMA Observatory is operated by ESO, AUI/NRAO and NAOJ.


\end{document}